\documentclass[11pt]{article}

\usepackage{amsfonts,amsmath,amssymb,graphics}
\newcommand{\CITE}{\cite}

\newcommand{\R}{\mathbb{R}}
\newcommand{\nE}{\widehat{\mathbf{E}}}
\newcommand{\nV}{\widehat{\mathbf{V}}}

\newcommand{\I}{\mathrm{i}}
\newcommand{\pr}{\mathbf{P}}
\newcommand{\ex}{\mathbf{E}}
\newcommand{\sh}{\mathbf{S}}
\newcommand{\V}{\mathbf{V}}
\newcommand{\VaR}{\mathrm{VaR}}

\newcommand{\notthis}[1]{}
\newcommand{\Real}{\mathrm{Re}\,}
\newcommand{\indic}[1]{\mathbf{1}[#1]}

\newcommand{\invinti}{\int_{\mathcal{I}}}
\newcommand{\invintr}{\int_{\mathcal{R}}}
\newcommand{\invintip}{\int_{\mathcal{I}+}}

\newcommand{\invintipm}{\int_{\mathcal{I}\pm}}
\newcommand{\invintrp}{\int_{\mathcal{R}+}}

\newcommand{\invintrpm}{\int_{\mathcal{R}\pm}}
\newcommand{\gtrlessdot}{{\,\raisebox{1mm}{$\gtrdot$}\!\!\!\!\raisebox{-1mm}{$\lessdot$}\,\,}}
\renewcommand{\sp}{\hat{s}}
\newcommand{\spz}{\hat{z}}
\newcommand{\half}{\frac{1}{2}}

\newcommand{\blobi}{\frac{1}{2\pi\I}}
\newcommand{\pbar}{\overline{p}}

\newcommand{\cdl}{\,|\,}
\newcommand{\seq}{\!=\!}
\newcommand{\shalf}{{\textstyle\frac{1}{2}}}
\newcommand{\deriv}[2]{\frac{\partial{#1}}{\partial{#2}}}
\newcommand{\dderiv}[3]{\frac{\partial^2{#1}}{\partial{#2}\partial{#3}}}
\newcommand{\Dderiv}[2]{\frac{\partial^2{#1}}{\partial{#2}^2}}

\begin{document}

\title{\bf Saddlepoint methods in portfolio theory}
\author{Richard J. Martin\footnote{Reproduced in \emph{Handbook of Credit Derivatives}, eds.\ A. Rennie \& A. Lipton, OUP, 2011}}
\maketitle

\begin{abstract}

We discuss the use of saddlepoint methods in the analysis of portfolios, with particular reference to credit portfolios. The objective is to proceed from a model of the loss distribution, given through probabilities, correlations and the like, to an analytical approximation of the distribution. Once this is done we show how to derive the so-called risk contributions which are the derivatives of risk measures, such as a given quantile (VaR) or expected shortfall, to the allocations in the underlying assets. These show, informally, where the risk is coming from, and also indicate how to go about optimising the portfolio.

\end{abstract}

%%%%%%%%%%%%%%%%%%%%%%%%%%%%%%%%%%%%%%%%%%%%%%%%%%%%%%%%%%%%%%%%%%%%%%%%%%%%%%%%%%%%%%%%%%%%%%%%%%%%%%%%%%%%%%%%%%%%%%%%%%%%%%%%%%%%%%%%%%%%%%%%%%%%
%%%%%%%%%%%%%%%%%%%%%%%%%%%%%%%%%%%%%%%%%%%%%%%%%%%%%%%%%%%%%%%%%%%%%%%%%%%%%%%%%%%%%%%%%%%%%%%%%%%%%%%%%%%%%%%%%%%%%%%%%%%%%%%%%%%%%%%%%%%%%%%%%%%%

\section{Introduction}

Problems in quantitative finance can, in the main, be put into one of two compartments: ascribing probabilities to events, and calculating expectations. The first class of problems is essentially what modelling is about, whether in derivatives pricing or in portfolio or risk management; the second is of course fundamental to their practical application. The distinction between the two compartments is worth making. On the one hand there are some computational techniques such as the binomial or trinomial tree that are applicable to many different models and asset classes (provided the underlying dynamics are diffusive). Conversely, there are some models that can be calculated using many different techniques, such as for example synthetic CDO models which can be tackled by numerical grid techniques, analytical approximations or Monte Carlo, or as a second example the emerging application of L\'evy processes in derivatives theory, which give rise to models that can be calculated using Fourier transforms, Laplace transforms, or Monte Carlo. Divorcing the model from its calculation is, therefore, quite a useful idea; as for one thing, when something goes wrong, as frequently happens in the credit world, one needs to know whether it is the model or the calculation of it that was at fault. In my experience, it is generally the former, as usually what has happened is that the modellers have overlooked an important source of risk and implicitly ascribed far too low a probability to it. That said, accurate and fast calculation is important, and that is what I shall be talking about here.

The construction of the distribution of losses, or of profit-and-loss, of a portfolio of assets is a well-established problem, whether in structuring multiasset derivatives, or running part of a trading operation, an investment portfolio or a bank. It can take many guises, according whether one is concerned with risk on a buy-and-hold view, or whether it is mark-to-market that is of more importance.

In this chapter we shall give the first complete exposition of the \emph{saddlepoint method} to the calculation and management of portfolio losses, in an environment that is quite general and therefore applicable to many asset classes and many models.
The methods described here apply equally well in the buy-and-hold and the mark-to-market contexts, and have been applied very successfully in both arenas.
Their most natural application is in credit, as through the collateralised debt obligation (CDO) market, and investment banks' exposure to bonds and loans, there has been an impetus to understand the losses that can be incurred by portfolios of credit assets. Portfolios of hedge fund exposures or derivatives of other types could also be analysed in the same framework. What makes credit particularly demanding of advanced analytics is that in general the losses that come from it are highly asymmetrical, with the long-credit investor generally receiving a small premium for bearing credit risk and occasionally suffering a very much larger loss when he chooses a `bad apple'. This asymmetry has also given rise to the exploration of risk measures other than the conventional standard deviation, with the Value at Risk (VaR) and expected shortfall (also known as conditional-VaR or CVaR). Another offshoot of the presence of credit portfolios has been the discussion of correlation in its various guises, and indeed in the form of CDO tranches one can actually trade correlation as an asset class. This has led to a standardisation of approach, with the conditional-independence approach now being almost universally used in the specification and implementation of portfolio models (the CDO world's brief flirtation with copulas being little more than another way of dressing up a fairly basic concept). The principle behind conditional independence is that the assets are independent conditionally on the outcome of some random variable often called a \emph{risk-factor}. By specifying the correlation this way, one has a clear interpretation of the behaviour of large-scale portfolios (which has led to the ASRF model of Gordy \CITE{Gordy03a}) and the joint distribution of all assets is specified reasonably parsimoniously.

The technique described here is called the saddlepoint approximation and comes from asymptotic analysis and complex variable theory. Although common in statistics and mathematical physics, its first application in portfolio theory seems to have been 1998 \CITE{Arvanitis98}, when the present author applied it to the calculation of distributions arising from simple credit loss models; since then the method has been applied and improved by various authors \CITE{Gordy02, Glasserman06, Huang07}.

This chapter naturally divides into two parts. First, we have the construction of the distribution of losses, which is an essential ingredient of valuing CDO tranches and calculating risk measures at different levels of confidence. Secondly, we want to understand where the risk is coming from, by which more formally we mean the sensitivity of risk to asset allocation, an idea that is fundamental to portfolio theory and risk management and is at the heart of the Capital Asset Pricing Model (CAPM). The two parts are closely connected, as we shall see that in obtaining the loss distribution, much information about the risk contributions is already apparent and needs only a few simple computations to extract it. In the theory of risk contributions, we shall identify some problems with the sensitivity of VaR to asset allocation, and show that, somewhat paradoxically, analytical \emph{approximations} to VaR contribution are actually more useful than the `correct' answer. We will also reestablish the fact that shortfall does not have these technical difficulties, and give a complete derivation of the sensitivity theory in both first and second order. The second-order behaviour will be shown to be `well-behaved' by which we mean that the saddlepoint approximation preserves the convexity and therefore is a reliable platform for portfolio optimisation (unlike VaR).

\section{Approximation of Loss Distributions}

\subsection{Characteristic functions}

Portfolio analytics can, in the main, be understood by reference to the characteristic function (CF) of the portfolio's distribution. The characteristic function of the random variable $Y$, which from now on will denote the loss (or P\&L) of the portfolio in question, is defined as the following function of the complex\footnote{As usual i denotes $\sqrt{-1}$.} variable $\omega$:
\[
C_Y(\omega) = \ex[e^{\I\omega Y}].
\]
The density of $Y$ can be recovered from it by the inverse Fourier integral
\begin{equation}
f_Y(x) = \frac{1}{2\pi} \int_{-\infty}^\infty C_Y(\omega) e^{-\I \omega x}\,d\omega.
\label{eq:dens_fourier}
\end{equation}
If the distribution of $Y$ is discrete then the convergence is delicate and the result has to be interpreted using delta-functions: $\frac{1}{2\pi}\int_{-\infty}^\infty e^{\I \omega (y-x)}\,d\omega=\delta(y-x)$. 

Obviously we have to be able to efficiently compute $C_Y$. The ingredients are: (i) the characteristic function of independent random variables is the product of their characteristic functions, and (ii) the characteristic function is an expectation, so one can do the usual trick of conditioning on a risk-factor and then integrating out. This construction is universal in portfolio problems, so we can review a few examples now.

\subsubsection*{Example: CreditRisk+}

Consider first a default/no-default model. Let the $j$th asset have a loss net of recovery $a_j$ and a conditional default probability $p_j(V)$, where $V$ is the risk-factor.
 Then
\[
C_Y(\omega) = \ex\left[\prod_j \big(1-p_j(V)+p_j(V)e^{\I \omega a_j}\big)\right]
\]
Make the assumption that the default probabilities are quite small, so that
\[
C_Y(\omega) \approx \ex\left[e^{\sum_j p_j(V)(e^{\I \omega a_j}-1)}\right],
\]
(the Poisson approximation), and then impose the following model of the conditional default probability: $p_j(V)=\overline{p}_j\cdot V$, so that the action of the risk-factor is to scale the default probabilities in proportion to each other. Of course the risk-factor has to be nonnegative with probability 1. Then
\[
C_Y(\omega) \approx \ex\left[e^{\sum_j \overline{p}_j(e^{\I \omega a_j}-1)V}\right] = M_V\!\left(\sum_j \overline{p}_j(e^{\I \omega a_j}-1)\right)
\]
where $M_V(s)=\ex[e^{sV}]$ denotes the moment-generating function (MGF) of $V$. It is then a matter of choosing a distribution for $V$ (apart from positivity, it needs to have a MGF that is known in closed form, so the Gamma and inverse Gaussian distributions are obvious candidates). Incidentally CreditRisk+ is configured to have exposures that are integer multiples of some quantum, and this enables the loss distribution to be obtained recursively\footnote{Which imposes further restrictions on the distribution of $V$; see \CITE{Lehrbass04} for a full discussion of the Panjer recursion.}. 

\subsubsection*{Example: Extended CreditRisk+; Gaussian copula}

Sometimes $C_{Y|V}$ is known in closed form but $C_Y$ is not. Here are two examples.

An objection to the basic CreditRisk+ model as just described is that assets can default more than once, which for low-grade assets is a problem. In that case we cannot use the Poisson approximation, and instead have to calculate the CF numerically. One therefore has
\begin{equation}
C_Y(\omega) = \int_{-\infty}^\infty \prod_j \big(1-p_j(v)+p_j(v)e^{\I \omega a_j}\big)f(v)\, dv
\label{eq:CFgen}
\end{equation}
with $p_j(v)=\min(\pbar_j\, v,1)$ and $f(v)$ denoting the density of $V$ (which is zero for $v<0$).

The Gaussian copula model requires the same treatment. In that case (\ref{eq:CFgen}) can still be used, but with different ingredients: $p_j(V) = \Phi\!\left(\frac{\Phi^{-1}(\overline{p}_j)-\beta_j V}{\sqrt{1-\beta_j^2}}\right)$ and $V\sim \mathrm{N}(0,1)$ for the risk-factor.

It is apparent that in this framework any model can be handled, once the distribution of $V$ and the `coupling function' $p_j(V)$ are specified. In fact, it is worth noting in passing that specifying a model this way leads to overspecification, as replacing $V$ by $h(V)$ (with $h$ some invertible function) and $p_j$ by $p_j\circ h^{-1}$ leaves the model unaffected. One can therefore standardise all one-factor models to have a Normally-distributed risk-factor, which simplifies the implementation somewhat.

\subsubsection*{Example: Merton model}

There are also situations in which even $C_{Y|V}$ is not known in closed form, though to find examples we have to go beyond the simple default/no-default model. A detailed exposition of mark-to-market (MTM) models would take us outside the scope of this chapter, but here is the basic idea. We adopt the Merton approach, in which the holder of debt is short a put option on the firm's assets. The put value relates to the firm value through the Black-Scholes put formula (for a simple model, at least), and the values of differerent firms' debt may be correlated by correlating their asset levels. If the firms' asset returns $Z_j$ are distributed as $\mathrm{N}(0,1)$ after suitable normalisation, then a simple Gaussian model is the obvious choice:
\[
Z_j = \beta_j V + \sqrt{1-\beta_j^2} \, U_j,
\]
where $V$ is the common part and $U_j$ the idiosyncratic part. Let the value of the debt be some function $g_j(Z_j)$ say. Then, conditioning on $V$ and integrating out the idiosyncratic return $U_j$, we have 
\[
C_{X_j|V}(\omega) = \int_{-\infty}^\infty \exp \!\left({\I\omega\, g_j\big(\beta_j V+\sqrt{1-\beta_j^2}\,u\big)}\right) \frac{e^{-u^2/2}}{\sqrt{2\pi}} \,du
\]
and the integral has to be done numerically. On top of that, the outer integral over $V$ has to be done.
This can be viewed as a continuum analogue of `credit migration' (the more common discrete version having been  discussed by Barco \CITE{Barco04}). The simple model above with Normal distributions is not to be recommended for serious use because the probability of large spread movements is too low, but one can build more complex models in which jump processes are used and from the point of view of risk aggregation the same principles apply.

\subsection{Inversion}

Having arrived at the characteristic function, we now wish to recover the density, tail probability or whatever. The density of the random variable $Y$ can be recovered from (\ref{eq:dens_fourier}) which can, for conditional independence models, be written in two equivalent forms:
\begin{eqnarray}
f_Y(x) &=& \frac{1}{2\pi} \int_{-\infty}^\infty \ex[C_{Y|V}(\omega)] e^{-\I \omega x}\,d\omega \label{eq:invdirect} \\
f_Y(x) &=& \ex\left[ \frac{1}{2\pi} \int_{-\infty}^\infty C_{Y|V}(\omega) e^{-\I \omega x}\,d\omega \right]
\label{eq:invindirect}
\end{eqnarray}
(i.e.\ the `outer integration' over $V$ can be done inside the inversion integral, or outside). Other expressions such as the tail probability and expected shortfall can be dealt with similarly.

In simple cases, such as default/no-default with equal loss amounts, the most direct route is to assemble the loss distribution on a grid (see e.g. \CITE{Burtschell05}). The distribution of independent losses is found recursively, building the portfolio up asset by asset, and for conditionally-independent losses one follows the usual ``condition, find distribution, integrate-out'' route: this is, in essence, eq.~(\ref{eq:invindirect}). When the loss amounts are not identical, this procedure can still be followed as long as one is prepared to bucket the losses so that they do end up on a grid. Note that for very large portfolios this method is very inefficient because the computation time is proportional to the square of the portfolio size. The method is also unsuited to continuous distributions.

For the aggregation of arbitrary distributions the Fast Fourier Transform, essentially a numerical integration of the Fourier transform, is a useful tool. As the inversion integral is a linear function of the characteristic function, it does not matter which of (\ref{eq:invdirect},\ref{eq:invindirect}) is followed, though the former requires fewer calls to the FFT routine and is therefore preferable. A few things should be borne in mind about FFTs: first, the FFT is a grid method and the use of distributions that are not precisely represented on a grid can cause artefacts; secondly, although the inversion is fast, the evaluation of the characteristic functions at each gridpoint is not always, and this is where most of the computation time is spent.

Finally we have analytical approximations  e.g.\ Central Limit Theorem (CLT), Edgeworth, Saddlepoint. In their classical form these are large portfolio approximations for sums of independent random variables, and they have an obvious advantage over numerical techniques on large portfolios: numerical methods spend an inordinately long time computing a distribution that the CLT would approximate very well knowing only the mean and variance (which are trivial to compute). 
These methods are nonlinear in the CF and so (\ref{eq:invdirect}) and (\ref{eq:invindirect}) give \emph{different} results.

When the analytical approximation is done inside the expectation, as in (\ref{eq:invindirect}), it is said to be \emph{indirect}. If outside, as in (\ref{eq:invdirect}), so that the unconditional CF is being handled, it is \emph{direct}.
It is argued in \CITE{Martin06a} that the indirect method is `safer', because it is still being applied to independent variables, with the outer integration having essentially no bearing on the accuracy of the approximation: by contrast, the direct method is more like a `black box'.
We concentrate on the indirect method that we shall be using throughout this chapter, with only a short section on the direct method. 
(In any case, the formulae for the direct methods will be obvious from the indirect ones.)

%%%%%%%%%%%%%%%%%%%%%%%%%%%%%%%%%%%%%%%%%%%%%%%%%%%%%%%%%%%%%%%%%%%%%%%%%%%%%%%%%%%%%%%%%%%%%%%%%%%%%%%%%%%%%
\subsection{Saddlepoint approximation: concepts, and approximation to density}

We need to introduce some more notation, in the shape of the moment-generating function (MGF), 
$M_Y(s)=\ex[e^{sY}]$, and when doing so we understand that it is to be evaluated for values of $s$ that might not be purely imaginary. As $M_Y(s)=C_Y(\omega)$ when $s=\I\omega$, and $C_Y(\omega)$ exists for all $\omega\in\R$ regardless of the distribution of $Y$, it must hold that $M_Y(s)$ exists for all pure imaginary $s$. However, when using the notation $M_Y$ we will take it as read that $M_Y(s)$ exists for values of $s$ off the imaginary axis (more precisely, in some band $s_-<\Real(s)<s_+$ with $s_-<0<s_+$. This imposes the restriction that the distribution of $Y$ decay at an appropriate rate in the tails (exponential decay is sufficient). Note that in many examples for credit risk (portfolios of bonds, CDS, CDOs etc.) the maximum loss is finite and so this condition is automatically satisfied. It is also satisfied for anything Normally distributed. The term cumulant-generating function (KGF) is used for $K_Y(s)=\log M_Y(s)$.

The most well-known results for the approximation of the distribution of a sum of independent random variables
\[
Y = \sum_{i=1}^n X_j,\qquad (X_j)\mbox{ i.i.d.},
\]
are the Central Limit (CLT) and Edgeworth expansions, but they are found not to approximate the tail well: they give best accuracy in the middle of the distribution. So a neat idea is to change measure by `tilting' so that the region of interest becomes near the mean. A convenient approach is to use an exponential multiplier:
\[
\tilde{f}_Y(y) = \frac{e^{\lambda y}f_Y(y)}{M_Y(\lambda)}
\]
where the denominator is chosen so as to make the `tilted' distribution integrate to unity.
It is easily verified that under the tilted measure ($\tilde{\pr}$) the mean and variance of $Y$ are $K_Y'(\lambda)$ and $K_Y''(\lambda)$, where $K_Y=\log M_Y$. By choosing $K_Y'(\lambda)=y$, we have in effect shifted the middle of the distribution to $y$. We then argue that if $Y$ is very roughly Normal under $\pr$, and hence also under $\tilde{\pr}$, then its probability density at its mean must be about $1/\sqrt{2\pi\tilde{\sigma}^2}$ where $\tilde{\sigma}^2$ is its variance under $\tilde{\pr}$. Then
\[
f_Y(y) = e^{K_Y(\lambda)-\lambda y} \tilde{f}_Y(y) \approx \frac{e^{K_Y(\lambda)-\lambda y}}{\sqrt{2\pi K_Y''(\lambda)}}
\]
which is the saddlepoint approximation to the density of $Y$.
The ``$1/\sqrt{2\pi\tilde{\sigma}^2}$'' approximation is a lot better than appearances might suggest. For example, with an exponential distribution (density $e^{-x}$) the true density at the mean is $e^{-1}$ and the approximation is $1/\sqrt{2\pi}$, which is about 8\% too high: this in spite of the fact the exponential and Normal distributions have very different shape. Note that, when applied to the saddlepoint approximation, this 8\% error is observed uniformly across the distribution (because an exponential distribution remains exponential under tilting). 

This method, known as the \emph{Esscher tilt}, is a recurring theme in more advanced work, particularly on risk contributions \CITE{Martin01d}, which we shall develop later in this chapter, and in importance sampling \CITE{Glasserman05} where it is used to steer the bulk of the sampling to the point of interest on the distribution (usually a long way out in the tail where there would otherwise be very few samples). One of the vital ingredients is the uniqueness of the tilting-factor. This follows from the convexity of $K_Y$, which in turn follows by consideration of the quadratic $q(t)=\ex[(1+tY)^2e^{sY}]$. As $q$ is nonnegative its discriminant (``$b^2-4ac$'') must be $\le0$, and so $\big(\ex[Ye^{sY}]\big)^2\le \ex[e^{sY}]\ex[Y^2e^{sY}]$, which on rearrangement gives $(\log M_Y)''\ge0$, as required. We shall encounter the same method of proving convexity later on, in conjunction with the expected shortfall risk measure.

\begin{figure}[h!]
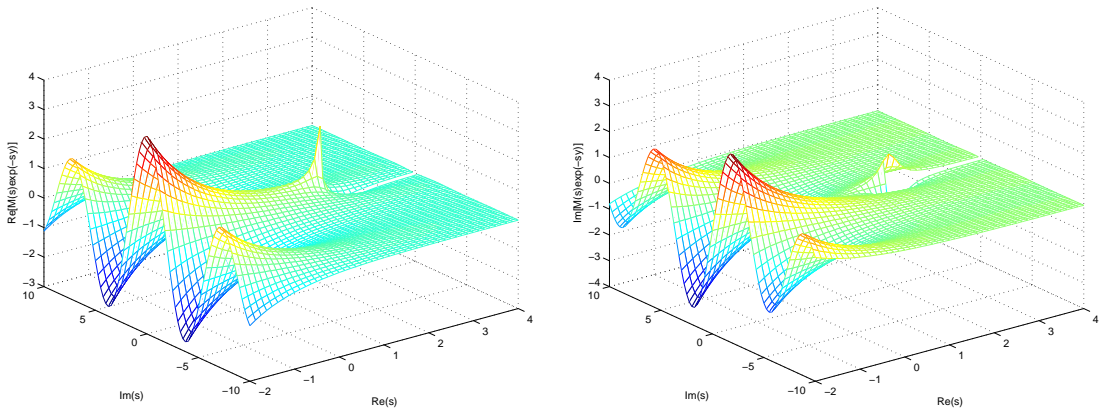

\begin{tabular}{cc}\scalebox{0.4}{\includegraphics*{wireplot1r.epsc}} &
\scalebox{0.4}{\includegraphics*{wireplot1i.epsc}} \end{tabular}
\caption{\small Real and imaginary parts of the inversion integrand $M(s)e^{-sy}$ where $M(s)=(1-\beta s)^{-\alpha}$ (the Gamma distribution), with $\alpha=1$, $\beta=0.5$, $y=1$. There is a branch point at $s=1/\beta$ and the plane is cut from there to $+\infty$. The integrand is oscillatory for contours parallel to the imaginary axis.}
\label{fig:wire1}
\end{figure}

\begin{figure}[h!]
\begin{tabular}{cc}\scalebox{0.4}{\includegraphics*{wireplot1a.epsc}} &
\scalebox{0.4}{\includegraphics*{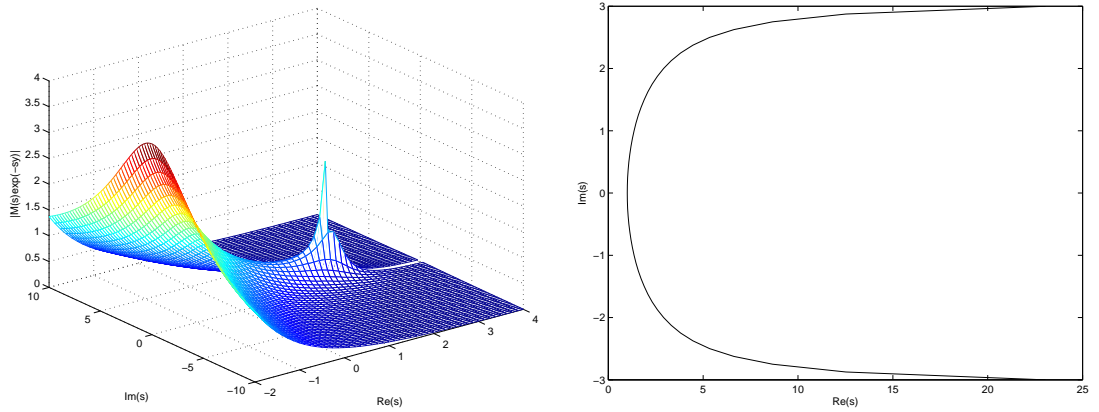}} \end{tabular}
\caption{\small Absolute value of the same inversion integrand $M(s)e^{-sy}$ as Figure~\ref{fig:wire1}, and path of steepest descent.
The contour runs in a horseshoe from $\infty-\pi\I$, through 1 (the saddlepoint), round to $\infty+\pi\I$.
}
\label{fig:wire2}
\end{figure}

Although the Esscher tilt is neat, our preferred approach uses contour integration, and explains where the term `saddlepoint' comes from. The reader is directed to \CITE{Bender78} for a fuller discussion of the asymptotic calculus. Assume first tthat $Y$ is the sum of independent and identically-distributed random variables, so that $K_Y=nK_X$ with $K_X(s)=\ex[e^{sX}]$. Expressing the density as a contour integral, distorting the path of integration $\mathcal{C}$  until it lies along the path of steepest descent---which is why the MGF must exist for $s$ off the imaginary axis---and then applying Watson's lemma gives:
\begin{eqnarray}
f_Y(y) &=& \blobi \int_{\mathcal{C}} e^{n(K_X(s)-sy/n)}\,ds \label{eq:spdens} \\
&\sim& \frac{e^{n(K_X(\sp)-\sp y/n)}}{\sqrt{2\pi nK_X''(\sp)}} \left[ 1+ \frac{1}{n} \left(\frac{K_X''''(\sp)}{8K_X''(\sp)^2} - \frac{5K_X'''(\sp)^2}{24K_X''(\sp)^3} \right) + O(n^{-2}) \right]\nonumber 
\end{eqnarray}
where $\sp$ is the value of $s$ that makes $K_Y(s)-sy$ stationary, that is, 
\begin{equation}
K_Y'(\sp)=y,
\label{eq:sp}
\end{equation}
from which we see that $\sp$ is interchangeable with $\lambda$.
We prefer this method of proof on the grounds that it is more upfront about the nature of the approximation, and is more useful when singular integrands such as the tail probability and shortfall are being examined. The term \emph{saddlepoint} refers to the nature of $K_Y(s)-sy$ at its stationary point: it has a local minimum when approached along the real axis, but a maximum when approached in the orthogonal direction, which is the path of integration here. A three-dimensional plot of the function would therefore reveal it to be in the shape of a saddle. By deforming the contour of integration so as to pass along the path of steepest descent, one avoids the problem of approximating a Fourier integral that is generally oscillatory  (Figures~\ref{fig:wire1},\ref{fig:wire2}).

A fundamental point is that the result can be written entirely in terms of $K_Y$ as
\begin{equation}
f_Y(y) \sim \frac{e^{K_Y(\sp)-\sp y}}{\sqrt{2\pi K_Y''(\sp)}} \left[ 1+ \left(\frac{K_Y''''(\sp)}{8K_Y''(\sp)^2} - \frac{5K_Y'''(\sp)^2}{24K_Y''(\sp)^3} \right) + O(K_Y^{-2}) \right].
\label{eq:spdensKY}
\end{equation}
Rather than being in descending powers of $n$, this approximation is in `descending powers of $K_Y$'.
One can therefore in principle use it in situations where $Y$ is not the sum of i.i.d.\ random variables. It can thus be applied to sums of variables that are independent \emph{but not identically distributed}, and this is one of the most important developments and insights in the theory and application. Common sense dictates, however, that the further away one gets from the non-identically distributed case, the less reliable the approximation is likely to be. For example, if there is a large exposure to one very non-Gaussian instrument in the portfolio, a poorer approximation should be expected (and indeed the correction term is considerable in that case).

\subsection{Tail probability}

The probability density function is not particularly useful in practice: the density function for a discrete model of losses consists of spikes and the fine structure of these is not especially important; one usually wants to know how likely it is for losses to exceed a certain level without having to integrate the density; and the approximated density does not exactly integrate to unity.
We therefore derive the tail probability and related quantities as integrals and approximate them.

The tail probability integral is in MGF form\footnote{$\mathcal{I}$ denotes the imaginary axis in an upward direction; the $+$ sign indicates that the contour is deformed to the right at the origin to avoid the singularity there (so it passes round the origin in an anticlockwise, or positive, direction).}
\begin{equation}
\pr[Y\gtrdot y] = \blobi \invintip e^{K_Y(s)-sy} \,\frac{ds}{s} .
\label{eq:tailpr}
\end{equation}
The symbol $\gtrdot$ is to be interpreted thus:
\[
\pr[Y\gtrdot y] \equiv \pr[Y>y] + \shalf \pr[Y=y].
\]
The reason for the $\half \pr[Y=y]$ term is that if $Y$ contains probability mass at $y$ then $M_Y(s)e^{-sy}$ contains a constant term of strength $\pr[Y=y]$, and now the interpretation of (\ref{eq:tailpr}) is rather delicate: the basic idea is to subtract out the divergent part of the integral and observe that the `principal value' of $\blobi\int_{-\I\infty}^{\I\infty} ds/s$ is $\half$. 
Notice therefore that if the true distribution of $Y$ is discrete then a naive comparison\footnote{As, for example, in a discussion by Merino \& Nyfeler \CITE[Ch.~17]{Lehrbass04}.} of $\pr[Y>y]$ with (\ref{eq:tailpr}) (or any continuum approximation to (\ref{eq:tailpr})) produces appreciable errors, but these have little to do with the quality of the analytical approximation because they arise from the continuity correction. In our examples, when we plot the tail probability, it will be $\pr[Y\gtrdot y]$ vs $y$.

Daniels \CITE{Daniels87} points out that although the two methods of deriving the saddlepoint expansion for the density (Edgeworth using Esscher tilt, or steepest descent using Watson's lemma) give the same result, they give different results for the tail probability. In essence this problem arises because of the singularity at $s=0$. 

The method of Lugannani \& Rice, which is preferable and will be used in several places in this chapter, removes the singularity by splitting out a singular part that can be integrated exactly. The saddlepoint expansion is then performed on the remaining portion. We refer to Daniels \CITE{Daniels87} for a fuller exposition, but here is the idea. It is convenient when performing the asymptotic approximation for the term in the exponential to be exactly quadratic rather than only approximately so. This can be effected by changing variable from $s$ to $z$ according to
\begin{eqnarray*}
\shalf (z-\spz)^2 &=& \big(K_Y(s) - sy\big) - \big(K_Y(\sp) - \sp y\big) \label{eq:stow1} \\
-\shalf \spz^2 &=& K_Y(\sp) - \sp y
\label{eq:stow2}
\end{eqnarray*}
which ensures that $s=0\Leftrightarrow z=0$ and also  $s=\sp\Leftrightarrow z=\spz$.  Then
the integral can be split into a singular part that can be done exactly and a regular part to which the usual saddlepoint treatment can be applied. The result is
\begin{equation}
\pr[Y\gtrdot y] \sim \Phi(-\spz) + \phi(\spz) \left\{ \frac{1}{\sp \sqrt{K_Y''(\sp)}} - \frac{1}{\spz} + \cdots \right\} ,
\label{eq:LR}
\end{equation}
the celebrated \emph{Lugannani-Rice formula}. As pointed out by Barndorff-Nielsen \CITE{Barndorff91}, this has the flavour of a Taylor series expansion of the function $\Phi(-\spz+\cdots)$, which does look more like a tail probability than (\ref{eq:LR}) does; after doing the necessary algebra the result is
\begin{equation}
\pr[Y\gtrdot y] \sim \Phi\!\left( -\spz + \frac{1}{\spz} \ln \frac{\spz}{\sp \sqrt{K_Y''(\sp)}} +\cdots\right).
\label{eq:BN}
\end{equation}
Note that both expressions are regular at $\sp=0$ but need careful numerical treatment near there: the preferred method is to expand in a Taylor series in $\sp$, and one obtains
\[
\pr[Y\gtrdot y] \sim \Phi\!\left( -\frac{K_Y'''(0)}{6K_Y''(0)^{3/2}}+\cdots\right).
\]

An interesting consequence of the Lugannani-Rice formula is that, in the saddlepoint approximation, 
\begin{equation}
\Dderiv{}{y} \ln \pr[Y\gtrdot y] < 0;
\label{eq:logconcave}
\end{equation}
we say that the resulting distribution is \emph{log-concave}. Hence the graph of $\pr[Y\gtrdot y]$ vs $y$ on a logarithmic scale, as we often plot, will always `bend downwards'. Not all distributions are log-concave; for example the exponential distribution clearly is, and more generally so is the Gamma distribution with shape parameter $\ge 1$, and the Normal distribution is too (by virtue of the inequality $\ex[(Z-x)\indic{Z>x}]>0$); but the Student-t is not. To prove the result, note from log-concavity of the Normal distribution that
\[
\spz \Phi(-\spz) - \phi(\spz) < 0.
\]
After some minor manipulations this can be recast as
%Now multiply both sides by $\sp/\spz$, and add the approximated density, $\frac{1}{\sqrt{K_Y''(\sp)}} \phi(\spz)$, to each side, and multiply through by $\frac{1}{\sqrt{K_Y''(\sp)}} \phi(\spz)$, to obtain
\[
\frac{\sp\phi(\spz)}{\sqrt{K_Y''(\sp)}}
\left( \Phi(-\spz) + \frac{\phi(\spz)}{\sp\sqrt{K_Y''(\sp)}}  - \frac{\phi(\spz)}{\spz}  \right) < \left(\frac{\phi(\spz)}{\sqrt{K_Y''(\sp)}}\right)^2.
\]
By (\ref{eq:LR})
this can be written, with $\pr[Y\gtrdot y]$ abbreviated to $P(y)$,
\[
P''(y)P(y) < P'(y)^2,
\]
from\footnote{Differentiation w.r.t.~$y$ slots in a factor of $(-\sp)$, so the first term on the LHS is $-1\times$ the derivative of the density.} which (\ref{eq:logconcave}) is immediate.

% end notthis

\subsection{Tranche payoffs and expected shortfall (ESF)}

We are now in a position to deal with the ESF, which is defined by $\sh^+[Y]=\ex[Y\cdl Y>y]$ or $\sh^-[Y]=\ex[Y\cdl Y<y]$ (according as $Y$ denotes portfolio loss or portfolio value) where $y$ is the VaR at the chosen tail probability\footnote{From now on we shall assume that the distributions in question are continuous, so we shall be less careful about distinguishing $>$ from $\gtrdot$. This avoids the complication of defining VaR and ESF for discontinuous distributions (see \CITE{Tasche02a} for the generalisation).}. 
The integral representation of the ESF is
\[
\ex[Y \indic{Y\gtrlessdot y}] = 
%\frac{1}{2\pi\I} \invintipm M_Y'(s)e^{-sy}\,\frac{ds}{s} = 
\frac{1}{2\pi\I} \invintipm K_Y'(s)e^{K_Y(s)-sy}\,\frac{ds}{s} .
\]
By following the methods that we applied to the tail probability, we write the singularity in the integrand as the sum of a regular part and an analytically tractable singular part near $s=0$:
\[
\frac{K'_Y(s)}{s} = \frac{\mu_Y}{s} + \frac{K'_Y(s)-\mu_Y}{s} + O(s)
\]
with $\mu_Y$ the mean of $Y$. The singular part then gives rise to the same integral as the tail probability, calculated earlier, and the regular part can be given the usual saddlepoint treatment. We have
\begin{equation}
\ex[Y \indic{Y\gtrlessdot y}] \sim \mu_Y \pr[Y\gtrlessdot y] \pm \frac{y-\mu_Y}{\sp} f_{Y}(y), 
\label{eq:ESF_sp}
\end{equation}
which is easily evaluated using the expressions for the density and the tail probability.
Division by the tail probability then gives the ESF. As usual the formula is exact if $Y$ is Normally distributed, as can be verified longhand.

There is a close link to tranche payoffs (calls and puts on the loss distribution), as follows.
When computing the payoff in a tranche it is necessary to find the difference between two call payoffs, where the call strikes (denoted $y$ in what follows) are the attachment and detachment points of the tranche. The `call' and `put' payoffs are $C^+_y\equiv \ex[(Y-y)^+]$ and $C^-_y \equiv \ex[(y-Y)^+]$ which have integral representation
\[
C^\pm_y = \blobi \invintipm M_Y(s)e^{-sy}\,\frac{ds}{s^2} \\
\]
(incidentally the put-call parity formula $C^+_y-C^-_y$ = $\mu_Y-y$ can be inferred by combining the two paths, which collapse to a loop round the origin, where the residue is $M_Y'(0)-y=\mu_Y-y$).
Integration by parts reduces the double pole to a single pole:
\[
C^\pm_y = \blobi \invintipm (K_Y'(s)-y) e^{K_Y(s)-sy} \, \frac{ds}{s},
\]
clearly a close relative of the ESF integral. (This can also be arrived at by noting that  $\ex[(Y-y)^+]=\ex\big[(Y-y)\indic{Y>y}\big]$, which is what we did above for ESF.)
Following the same route as with ESF we obtain
\[
C^\pm_y \sim (\mu-y) \pr[Y \gtrlessdot y]+ \frac{y-\mu}{\sp} f_Y(y).
\]

\subsection{Examples 1 (Independence)}

\begin{figure}[h!]
\begin{tabular}{cc}\scalebox{0.6}{\includegraphics*{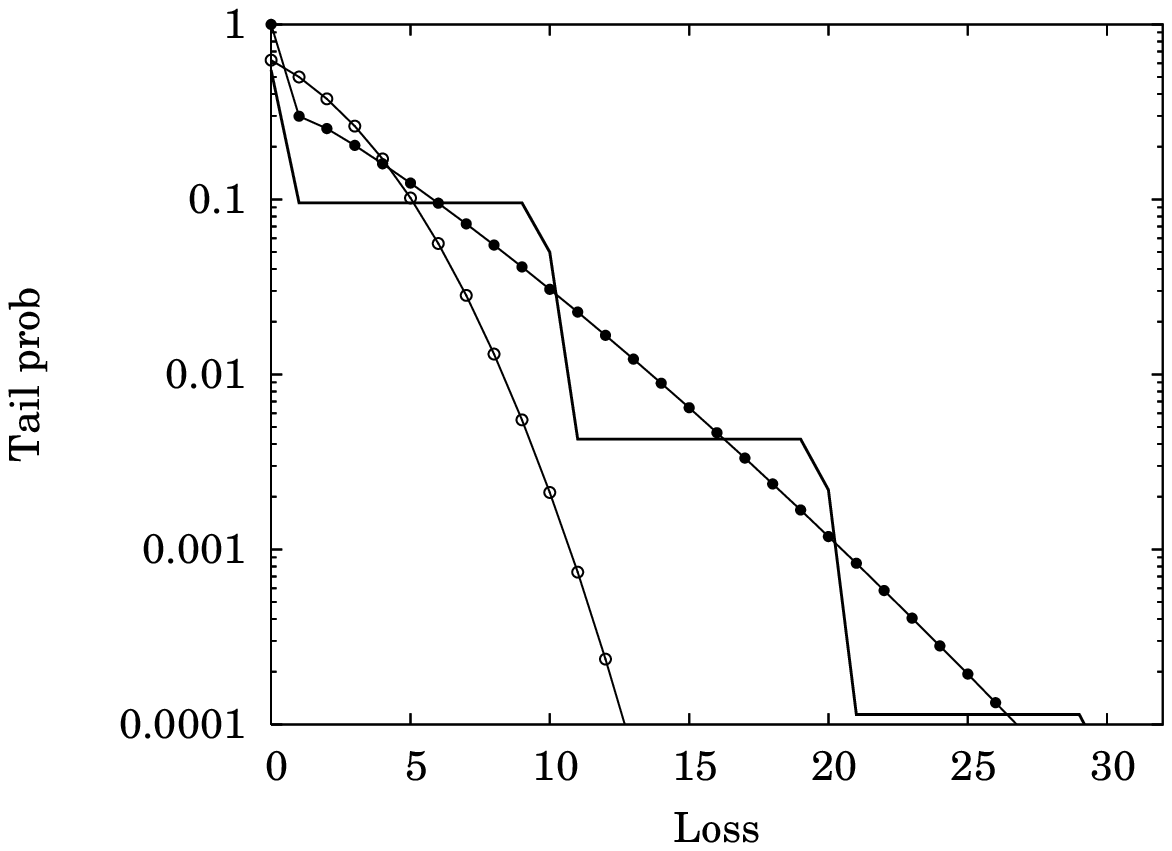}} &
\scalebox{0.6}{\includegraphics*{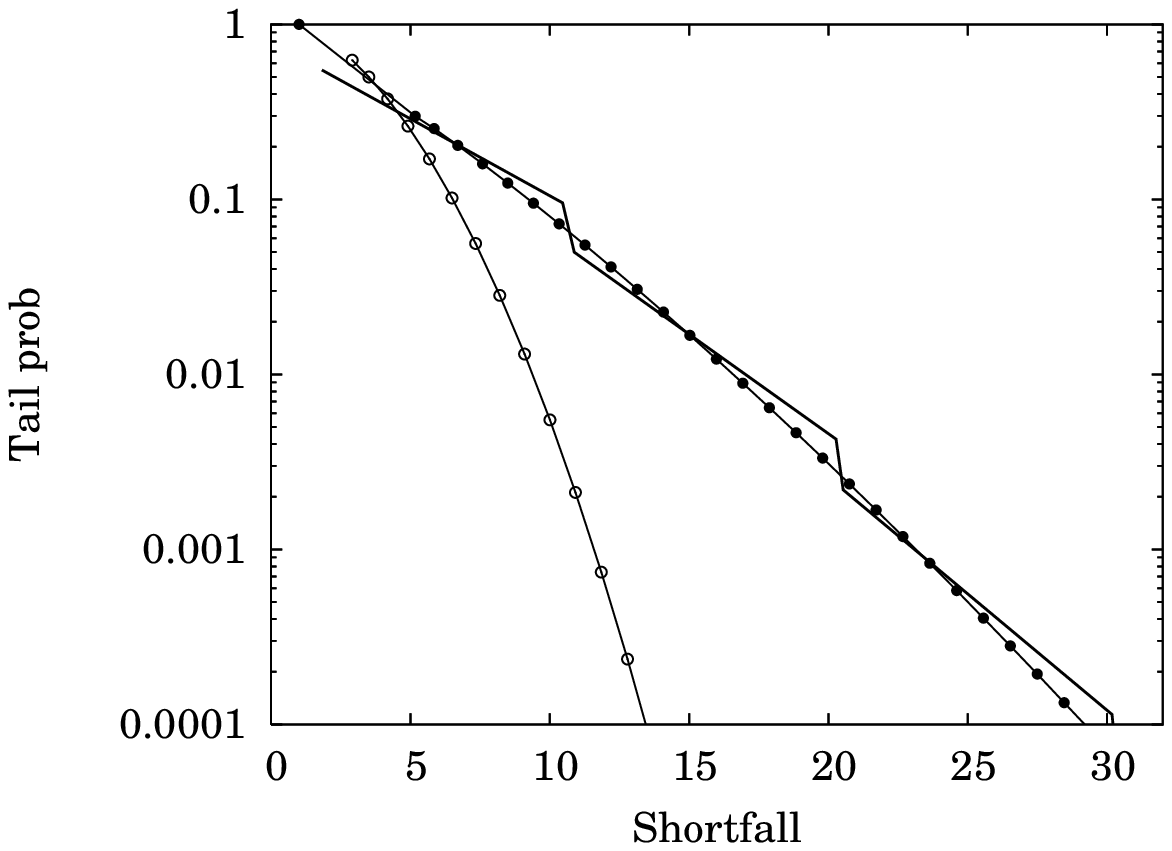}} \end{tabular}
\caption{\small Equal exposures, very small portfolio (10 assets). {$\bullet$}=Saddlepoint, {$\circ$}=CLT. Exact result (which is `steppy')  shown by unmarked line.}
\label{fig:pftest1}
\end{figure}

\begin{figure}[h!]
\begin{tabular}{cc}\scalebox{0.6}{\includegraphics*{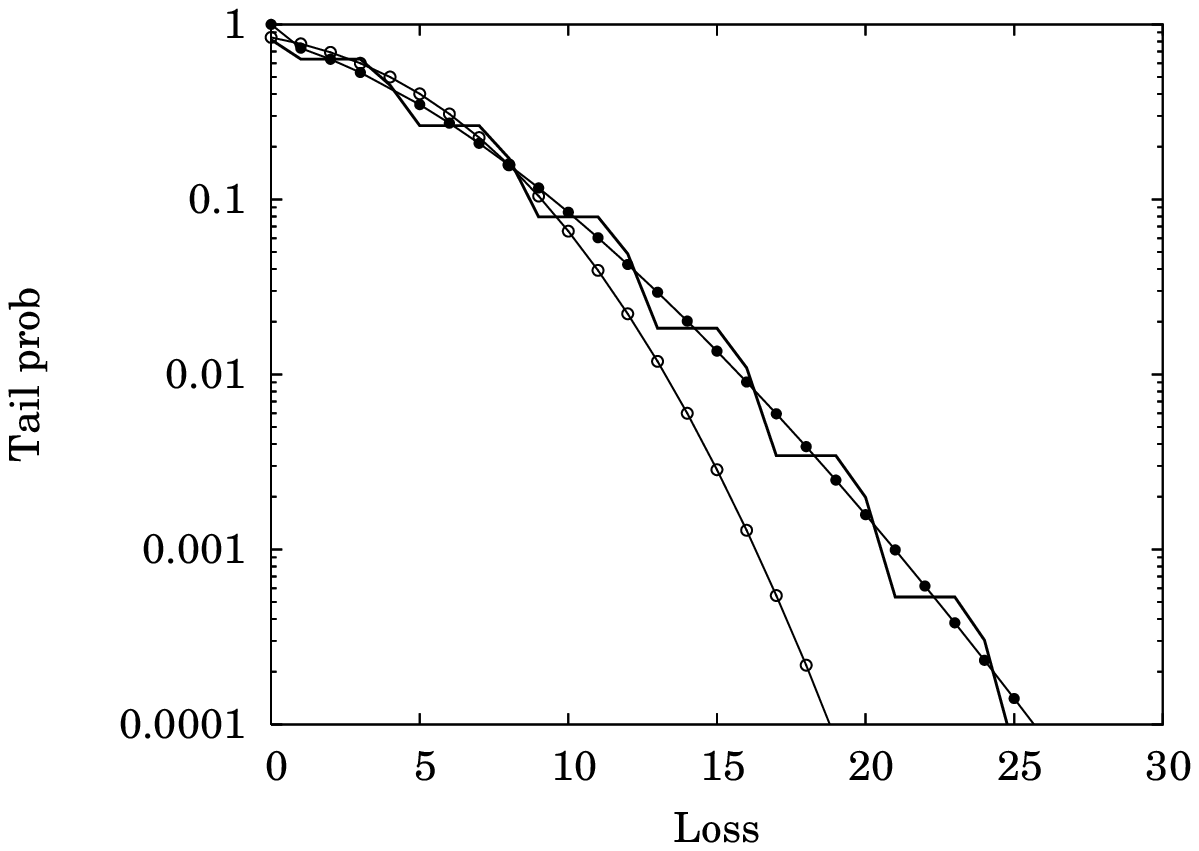}} &
\scalebox{0.6}{\includegraphics*{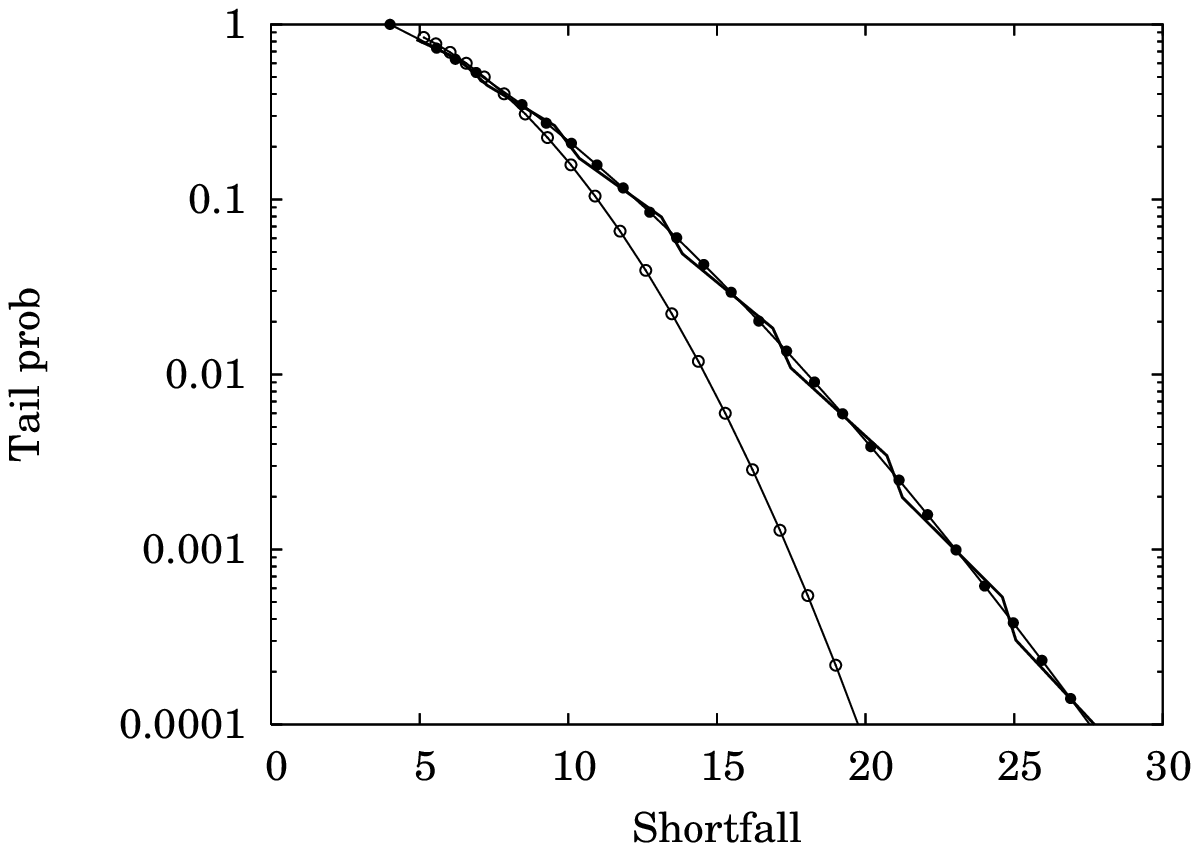}} \end{tabular}
\caption{\small As above but for larger portfolio (100 assets).}
\label{fig:pftest2}
\end{figure}

\begin{figure}[h!]
\begin{tabular}{cc}\scalebox{0.6}{\includegraphics*{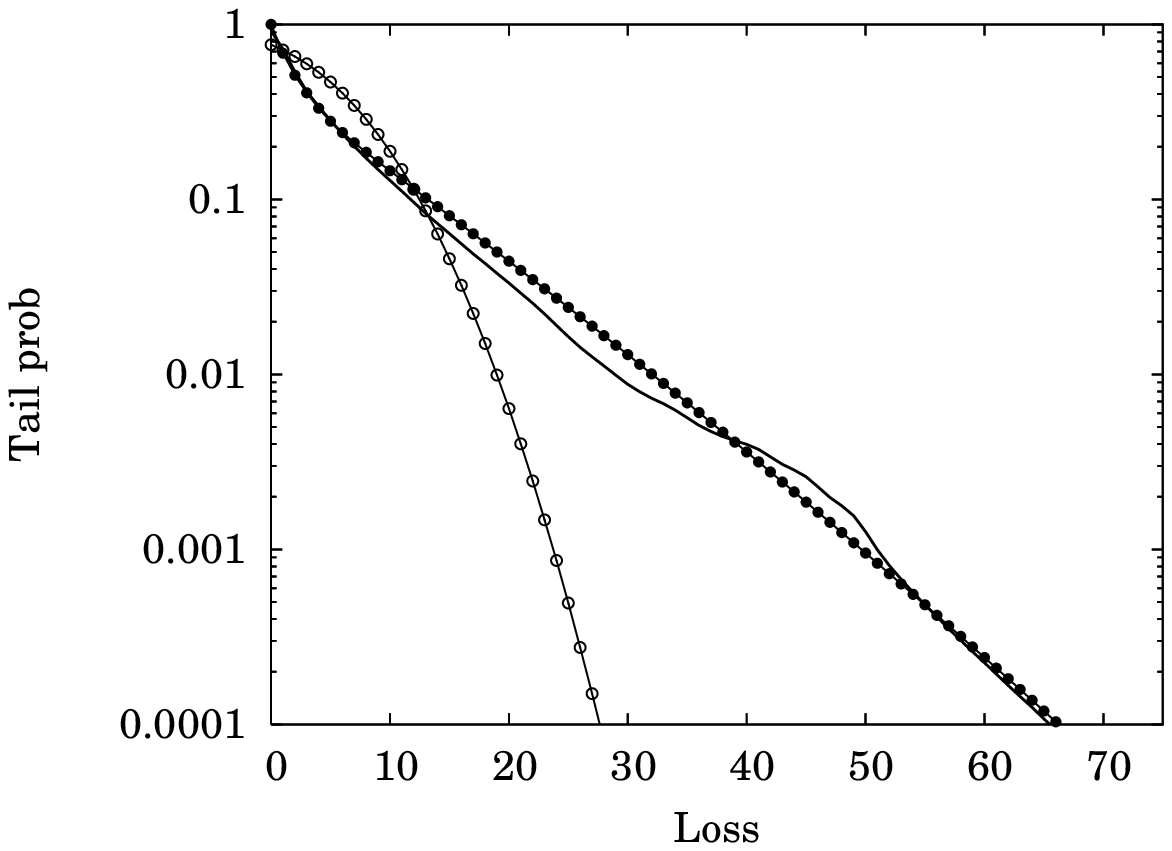}} &
\scalebox{0.6}{\includegraphics*{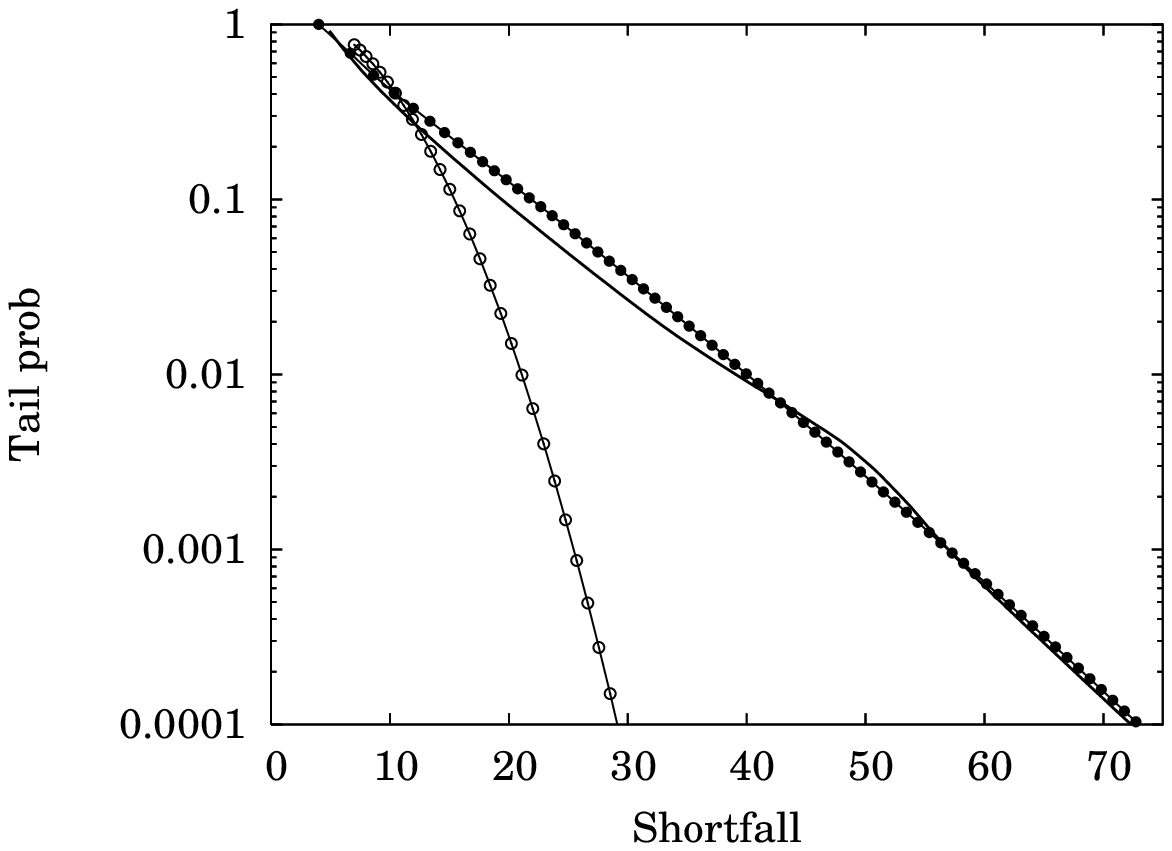}} \end{tabular}
\caption{\small Inhomogeneous portfolio (100 assets). The largest exposure is 10$\times$ median.}
\label{fig:pftest3}
\end{figure}

\begin{figure}[h!]
\begin{tabular}{cc}\scalebox{0.6}{\includegraphics*{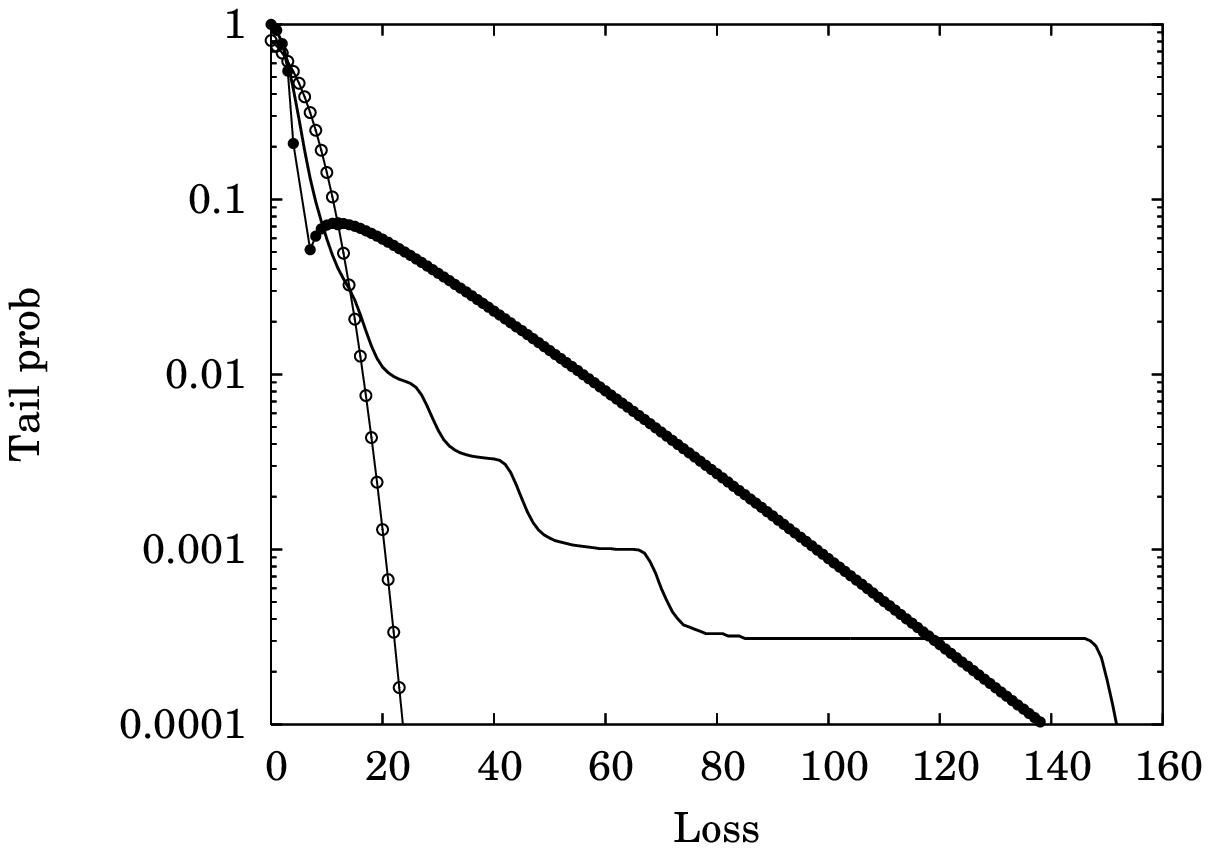}} &
\scalebox{0.6}{\includegraphics*{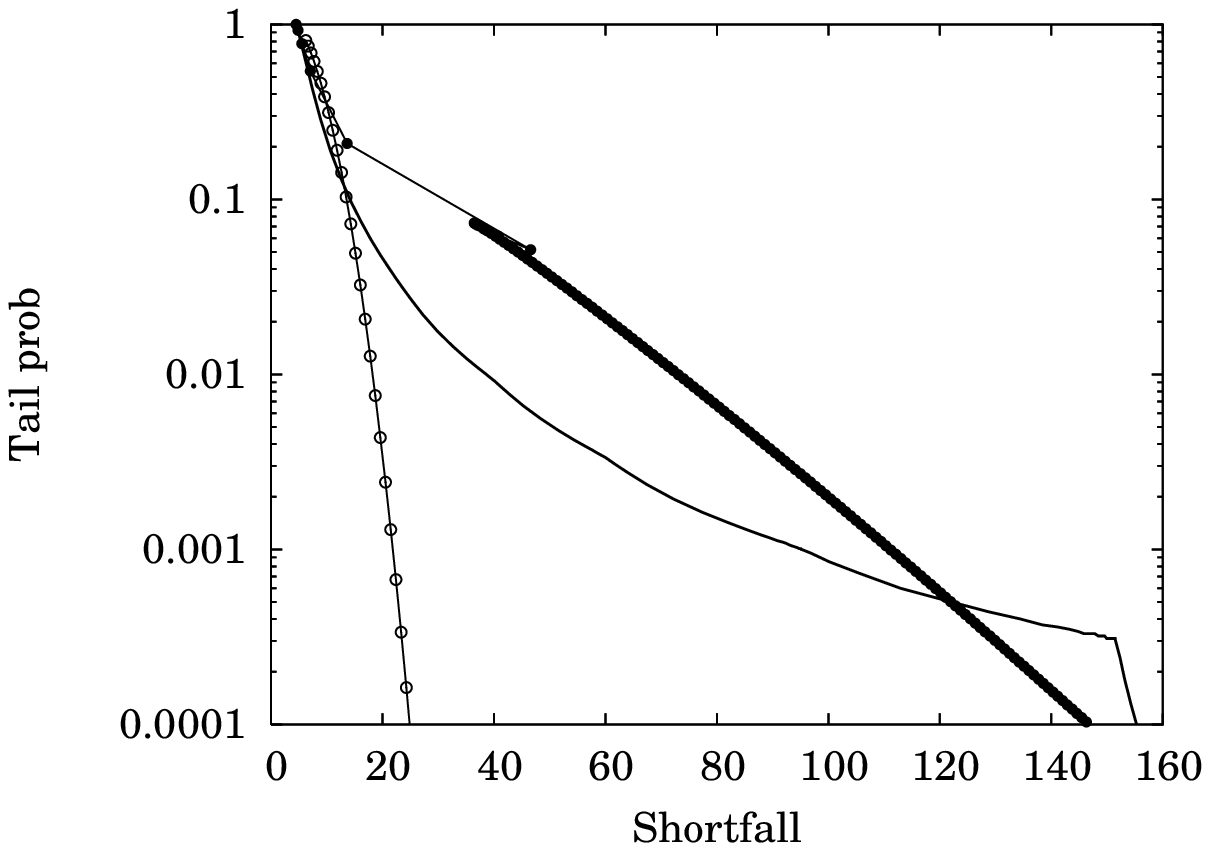}} \end{tabular}
\caption{\small An unlikely case: Extreme heterogeneity in which largest exposure is 150$\times$ median. The true distribution has a step in it at loss=150, which the saddlepoint approximation attempts to smooth out, thereby overestimating risk at lower levels.}
\label{fig:pftest4}
\end{figure}

It is now time for some numerical examples. We take a default/no-default model for ease of exposition. In each case we obtain the true distribution by inverting the Fourier integral numerically (by the FFT \CITE{NRC}), the saddlepoint approximation and the Central Limit Theorem.  This allows the analytical approximations to be verified. Different portfolio sizes and different default probabilities are assumed, as follows.
\begin{itemize}
\item
Figure~\ref{fig:pftest1}: 10 assets, exposure 10, default probability 1\%. Apart from the obvious problem of trying to approximate a discrete distribution with a continuous one, the saddlepoint accuracy is reasonably uniform. across the distribution.
\item
Figure~\ref{fig:pftest2}: 100 assets, exposure 4, default probability 1\%. This portfolio is more fine-grained, but note that even with 1000 assets the CLT is appreciably in error for higher quantiles.
\item
Figure~\ref{fig:pftest3}: 100 assets, unequal exposures (median 5, highest 50), default probabilities variable (typically 0.1--4\%; lower for the assets with higher exposure). Again the saddlepoint approximation works well and accuracy is roughly uniform across the distribution.
\item
Figure~\ref{fig:pftest4}: 100 assets, a couple of very large exposures (median 1, highest 150), default probabilities variable (typically 0.04--4\%; lower for the assets with higher exposure).
\end{itemize}

The last case is extreme and has a few features that make it difficult to deal with: a large step in the loss distribution caused by the very binary nature of the portfolio (if the biggest asset defaults, a huge loss is incurred; otherwise the losses are of little importance). In mark-to-market models there is a continuum of possible losses so this situation does not arise. Furthermore, even if a situation like this does crop up, the approximation is erring on the side of being conservative about the risk. Given that the period of the current `Credit Crunch' has witnessed many allegedly `very low probability' events, this form of approximation error is unlikely to result in bad risk management decisions.
In the context of a portfolio optimisation, the method would immediately start by chopping back such large exposures---and once it had done that, the approximation error would decrease anyway.

Later on (Figure~\ref{fig:rc2b}) in discussing risk contributions we shall give another example of a smallish inhomogeneous portfolio, showing that the approximation works well.
Other examples are given in \CITE{Martin01b, CSFB04}, the latter showing the results of applying the method to assets that are Gamma distributed (a continuous distribution).

\subsection{Accuracy and asymptoticity of saddlepoint approximation}

To understand why the saddlepoint approximation is accurate it is necessary first to understand in what sense the asymptotic expansion is being taken: the method is asymptotic for a large number of independent random variables being added, though as we have just seen, the number need not be very large in practice. Typically the relative  accuracy \emph{when the method is used only for addition of independent random variables} is roughly uniform across the whole distribution \CITE{Kolassa94} rather than just being a `tail approximation'.

For small portfolios, one can easily develop direct computations for establishing the loss distribution, essentially by considering all possible combinations of events. Asymptotic expansions work for portfolios that have a large number of assets. Why do we regard asymptotic methods as more useful? Essentially because there are more large numbers than there are small ones!

We have said that a `large number' of random variables have to be added, and to understand what `large' means in practice we can look at the first correction term (\ref{eq:spdens}). Clearly the smaller the higher cumulants the smaller the error. However, certain non-Gaussian distributions give very small errors: for example the error vanishes altogether for the inverse Gaussian distribution, while for the Gamma distribution with shape parameter $\alpha$ it is $-1/(12n\alpha)$ which is small when the shape parameter of $Y$, which is $n\alpha$, is large. (If $Y$ is exponentially distributed then the correction term can be calculated easily from (\ref{eq:spdensKY}) and it works out as $-\frac{1}{12}$, for all $s$; this almost exactly cancels the aforementioned $8\%$ error in the leading-order term. Jensen \CITE{Jensen95} has a more complete discussion.)

\notthis{
As another example, consider the Poisson distribution with mean $\mu$. It is nor difficult to verify that the probability density in the saddlepoint approximation is
\[
f_\mathrm{\sp}(y) = \frac{e^{-\mu}\mu^y}{\sqrt{2\pi y}(y/e)^y} \quad \mbox { c.f. } \quad \frac{e^{-\mu}\mu^y}{y!}; 
\]
in effect we have replaced $y!$ (in the exact, discrete formula) by Stirling's approximation to it, and this is known to be a good approximation even for $\mu$ as low as 1.
} % end notthis

Being more adventurous, one could even forget about independence and then use the saddlepoint approximation as a black box, as suggested in \CITE{Martin01b}. This is the `direct approach'; we shall return to it later.

\subsection{Conditionally-independent variables}

By inserting the outer integration over the risk-factor we obtain all the results for conditionally-independent variables.
This means that $\sp$, $\mu$ and the density and tail probability  are all factor-dependent. 
Treating things this way, we divorce the question of the saddlepoint approximation's accuracy from the choice of correlation model, as the approximation is only being applied to the distribution of a sum of \emph{independent} variables. This is important, because the subject of correlation is a vast one and one could never hope to check the method's accuracy for `all possible correlation models'. Making the choice of correlation model irrelevant is therefore a useful step.

For the density, one has
\begin{equation}
f_Y(y) \approx \ex\left[ \frac{e^{K_{Y|V}(\sp_V)-\sp_V y}}{\sqrt{2\pi K_{Y|V}''(\sp_V)}} \left( 1+ \left(\frac{K_{Y|V}''''(\sp_V)}{8K_{Y|V}''(\sp_V)^2} - \frac{5K_{Y|V}'''(\sp_V)^2}{24K_{Y|V}''(\sp_V)^3} \right) + \cdots\right)\right],
\label{eq:spdens_indir}
\end{equation}
and for the tail probability,  
\begin{equation}
\pr[Y\gtrdot y] \approx \ex\left[\Phi\!\left( -\spz_V + \frac{1}{\spz_V} \ln \frac{\spz_V}{\sp_V \sqrt{K_{Y|V}''(\sp_V)}} +\cdots\right)\right],
\label{eq:BN_indir}
\end{equation}
while for the shortfall,
\begin{equation}
\sh^\pm_y[Y] \approx \frac{1}{P^\pm} \ex \!\left[ \mu_{Y|V} \pr[Y \gtrlessdot y\cdl V]\pm \frac{y-\mu_{Y|V}}{\sp_V} f_{Y|V}(y) \right]
\label{eq:ESF_sp_indir}
\end{equation}
with $P^+,P^-$ the upper and lower tail probability (subscripts $V$ and $|V$ denote the dependence on $V$).
Incidentally the Central Limit Theorem would give the following, which is obtained by taking $\sp_V\to 0$:
\begin{equation}
\sh^\pm_y[Y] =  \frac{1}{P^\pm} \ex \!\left[ \mu_{Y|V} \Phi\!\left(\frac{\mu_{Y|V}-y}{\sigma_{Y|V}}\right) \pm \sigma_{Y|V} \phi\!\left(\frac{\mu_{Y|V}-y}{\sigma_{Y|V}}\right) \right],
\label{eq:esf-clt}
\end{equation}
with $\mu_{Y|V}$ and $\sigma_{Y|V}^2$ denoting the mean and variance of the portfolio conditional on the risk-factor; this can be obtained directly.

It is worth mentioning at this point the \emph{granularity adjustment} which gives a formula for VaR when a \emph{small} amount of unsystematic risk is added to the portfolio. This can be derived from the above equations.
Now if we wish to incorporate the effects of unsystematic risk we can model the loss as $Y=Y_\infty+U$,
with $Y_\infty=\ex[Y\cdl V]$ (the loss of the putative `infinitely granular portfolio', which need not exist in reality) and $U$ denoting an independent Gaussian residual of variance $\sigma^2$ which can depend on $Y_\infty$. The difference between the upper $P$-quantiles of $Y_\infty$ and $Y$ is given by the granularity adjustment (GA) formula (\CITE{Martin02a} and references therein):
\begin{equation}
\VaR_P[Y] \sim \VaR_P[Y_\infty] - \left. \frac{1}{2f(x)} \frac{d}{dx} \sigma^2(x)f(x) \right|_{x=\VaR_P[Y_\infty]},
\label{eq:GA_VaR}
\end{equation}
where $f$ is the density of $Y_\infty$.
The shortfall-GA is
\begin{equation}
\sh_P[Y] \sim \sh_P[Y_\infty] + \left. \frac{1}{2P} \sigma^2(x)f(x) \right|_{x=\VaR_P[Y_\infty]}.
\end{equation}
Note that the correction to shortfall is always positive (and analytically neater), whereas the correction to VaR is not; we will be discussing this issue in more detail later, and essentially it follows from non-coherence of the VaR risk measure \CITE{Artzner99,Tasche02a}.

%Likewise the call payoff becomes
%\[
%C^+_y \sim (\mu_{Y|V}-y) \pr[Y\gtrdot y\cdl V]+ \frac{y-\mu_{Y|V}}{\sp_V} f_{Y|V}(y).
%\]

The equation (\ref{eq:ESF_sp_indir}) and its simpler Central Limit analogue are the sum of two pieces and it is attractive to interpret them as, respectively, the contributions of systematic and specific risk to the portfolio ESF. This is because the first term is related to the variation of the conditional mean of the portfolio ($\mu_{Y|V}$) with the risk-factor and the second term is related to the residual variance ($\sigma^2_{Y|V}$) not explained by the risk-factor. Roughly, the first term is proportional to the interasset correlations (R-squared in KMV terminology or $\beta^2$ in CAPM) and the second to the reciprocal of the portfolio size. It turns out that this decomposition is analogous to the well-known result for the standard deviation,
\[
\V[Y] = \V\big[\ex[Y\cdl V]\big] + \ex\big[\V[Y\cdl V]\big] .
\]
This and related issues are discussed in \CITE{Martin07a}.

Note that the log-concavity property derived earlier is destroyed by the mixing operation, so in the conditionally-independent case it is only the $V$-\emph{conditional} distributions that are log-concave; the unconditional distribution may well not be.

\subsection{Computational issues}

It is worth mentioning some of the computational issues needed to implement a workable calculator.
A large proportion of the computation time is spent in the root-searching for $\sp$: the equation $K'_{Y|V}(s_V)=y$ has to be solved for each $V$, and if one is finding the VaR for some given tail probability, the loss level ($y$) will have to be adjusted in some outer loop. Consequently, it is worth devoting some effort to optimising the root-searching.
%When it comes to implementing these calculations, one needs a routine that evaluates $K_{Y|V}(s)$ and its derivatives for any $V$ and $s$. If one only takes the leading-order term, only two derivatives are strictly necessary. However, another routine is needed to calculate $\sp_V$ (from $K'_{Y|V}(s_V)=y$ and this has to be done by root-searching. If Newton-Raphson is used, only one derivative of $K'$ is needed, but higher-order analogues work better.
The first thing to note is that the saddlepoint is roughly given by $\sp\approx(y-K'(0))/K''(0)$, as a development of $K(s)$ around $s=0$ gives $K(s)=K'(0)s+\half K''(0)s^2+\cdots$ (which would be exact for a Normal distribution). The value $K''(0)^{-1/2}$, which is the reciprocal of the standard deviation, is a convenient `unit of measurement' that tells us how rapidly things vary in the $s$-plane: note that $s$ has dimensions reciprocal to $Y$, so if typical losses are of the order $10^6$ USD, then typical values of $\sp$ will be of order $10^{-6}\mbox{USD}^{-1}$. Secondly, as $K'$ is an analytic function, it makes sense to solve $K'(s)=y$ by Newton-Raphson, but higher-order variants (i.e.\ formulas that take into account the convexity of $K'$) generally converge much more quickly and have a wider basin of convergence. This means that $K'''(s)$ will need to be known for any $s$. It follows that the routine that evaluates $K_{Y|V}(s)$ had better calculate \emph{three} $s$-derivatives. (Recall also that $K'''(0)$ is explicitly used in a special case of the tail probability calculation when $\sp=0$.) 
A well-optimised root-searching routine should be able to find $\sp$ to machine precision in about three trials.

\subsection{Examples 2 (Conditional independence)}

\begin{figure}[h!]
\begin{tabular}{cc}\scalebox{0.6}{\includegraphics*{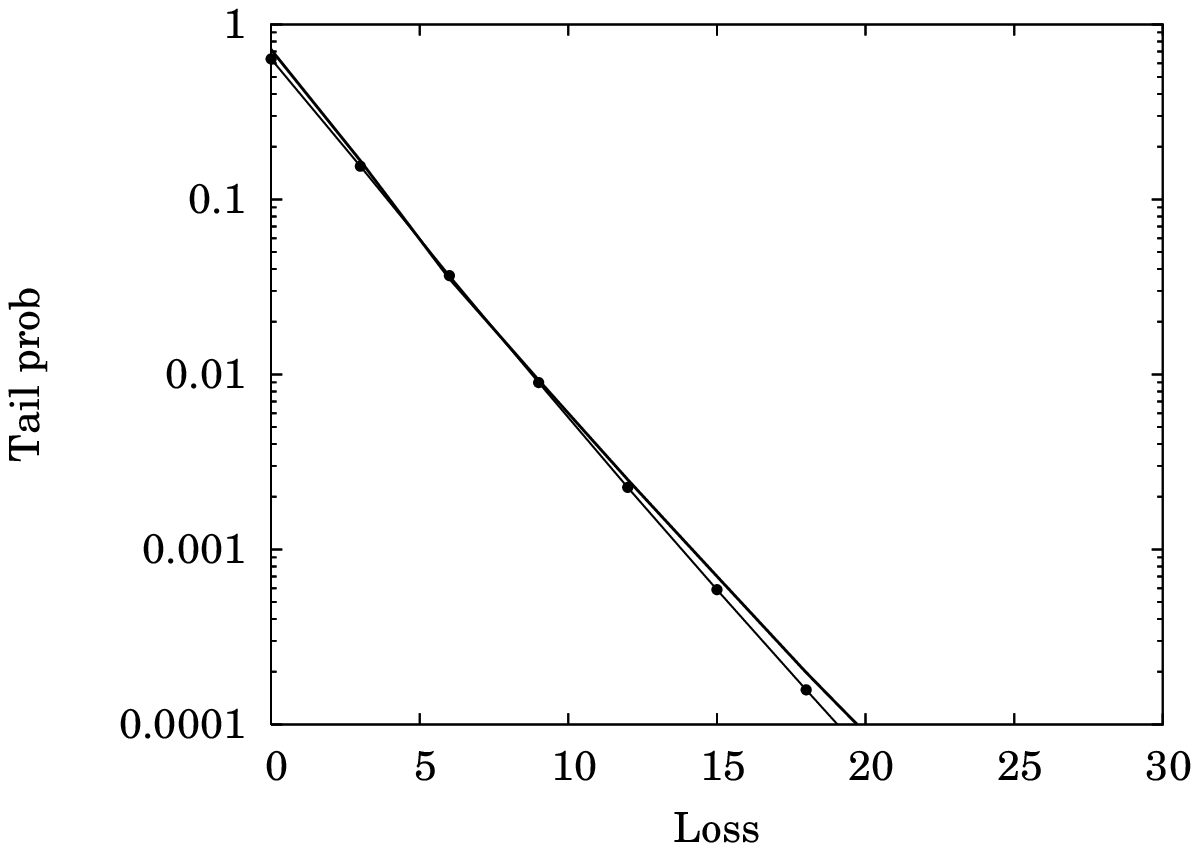}} &
\scalebox{0.6}{\includegraphics*{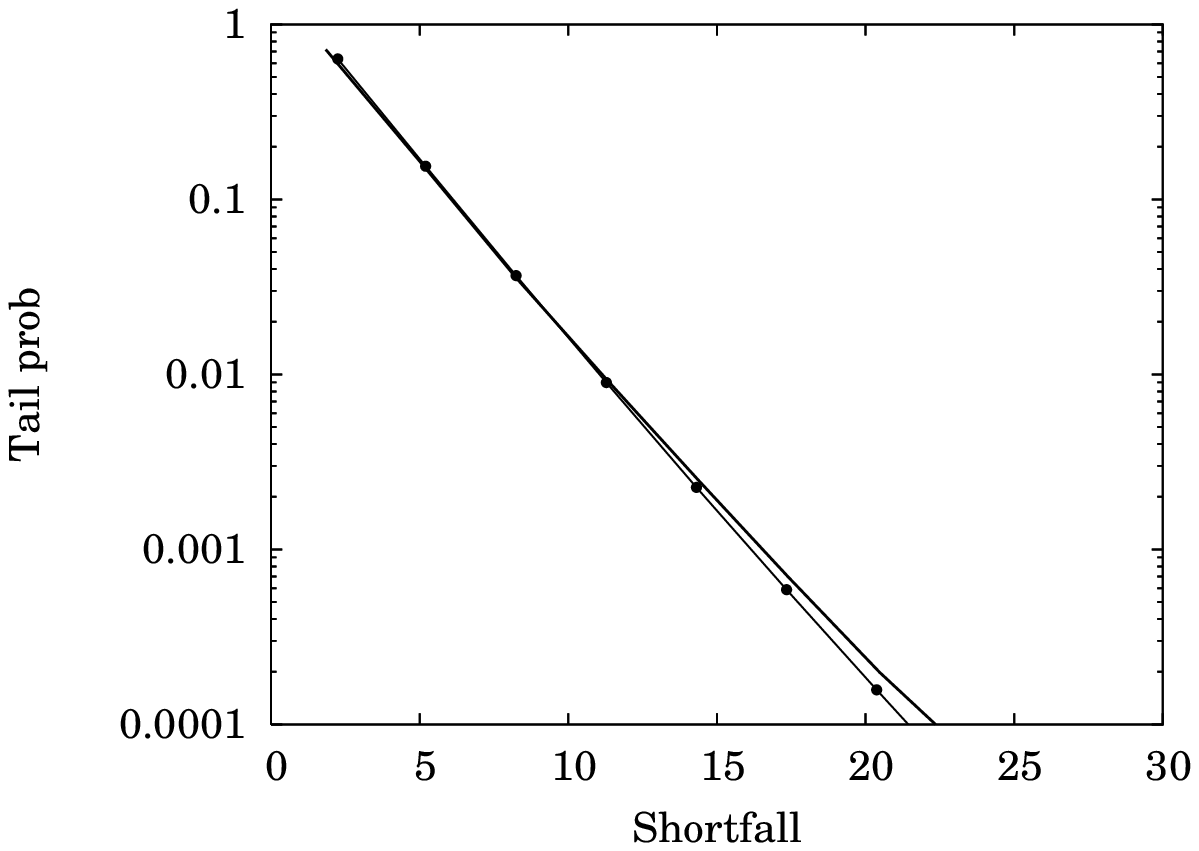}} \end{tabular}
\caption{\small Correlated test portfolio in Example 2: $\beta=0.3$.}
\label{fig:pftestc1}
\end{figure}

\begin{figure}[h!]
\begin{tabular}{cc}\scalebox{0.6}{\includegraphics*{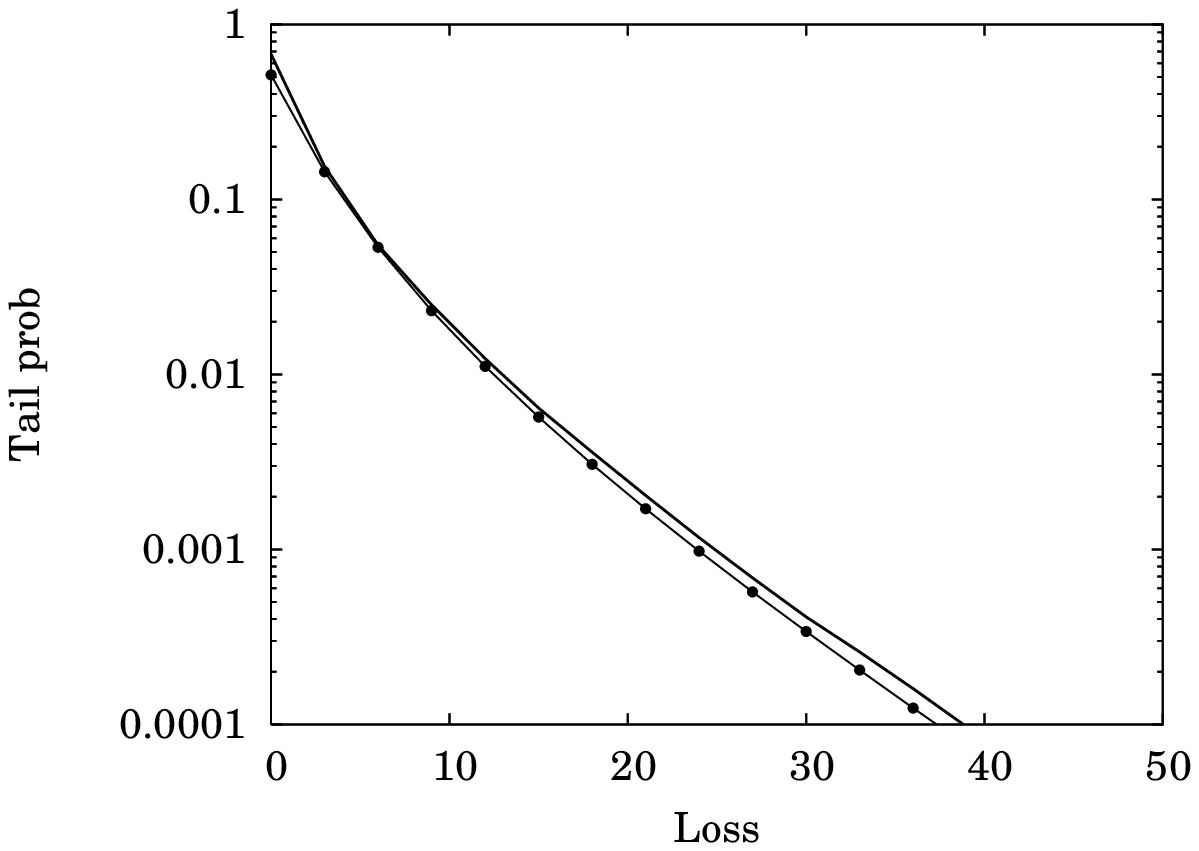}} &
\scalebox{0.6}{\includegraphics*{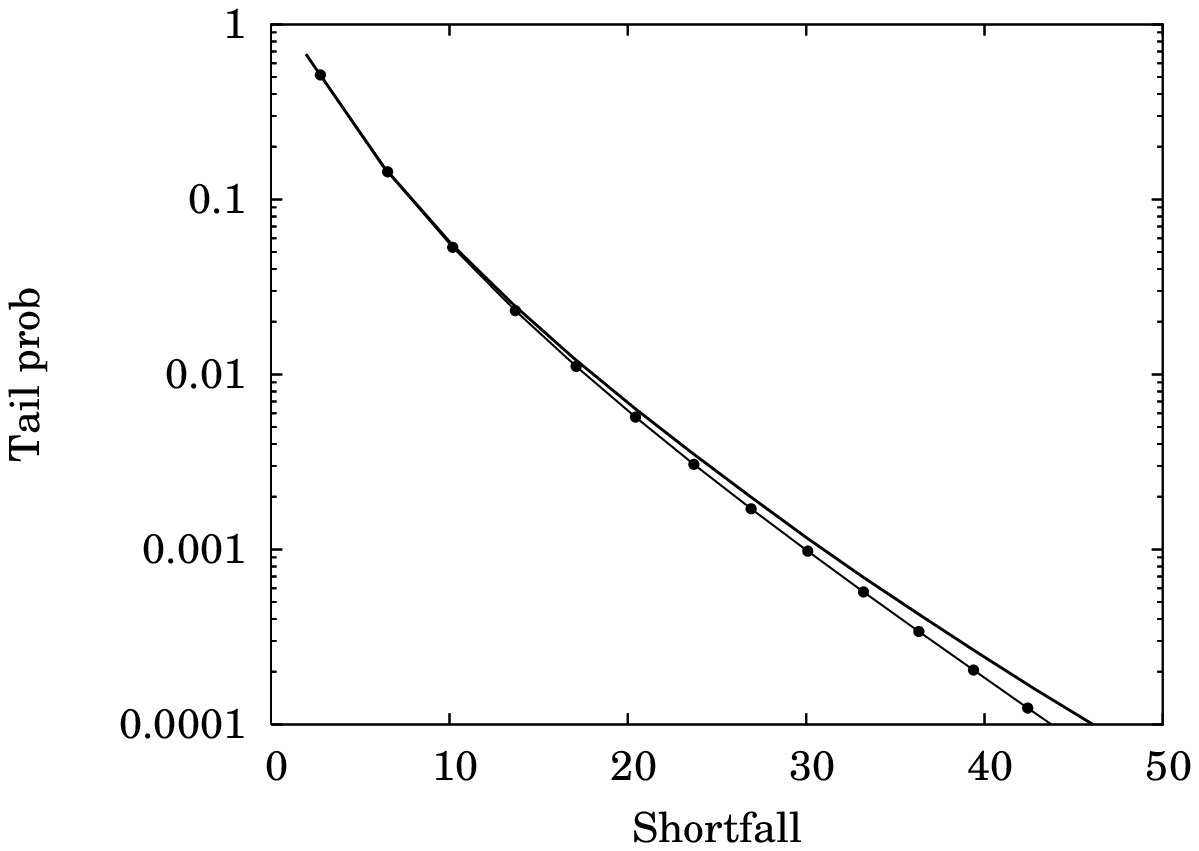}} \end{tabular}
\caption{\small As above but with $\beta=0.5$.}
\label{fig:pftestc2}
\end{figure}

\begin{figure}[h!]
\begin{tabular}{cc}\scalebox{0.6}{\includegraphics*{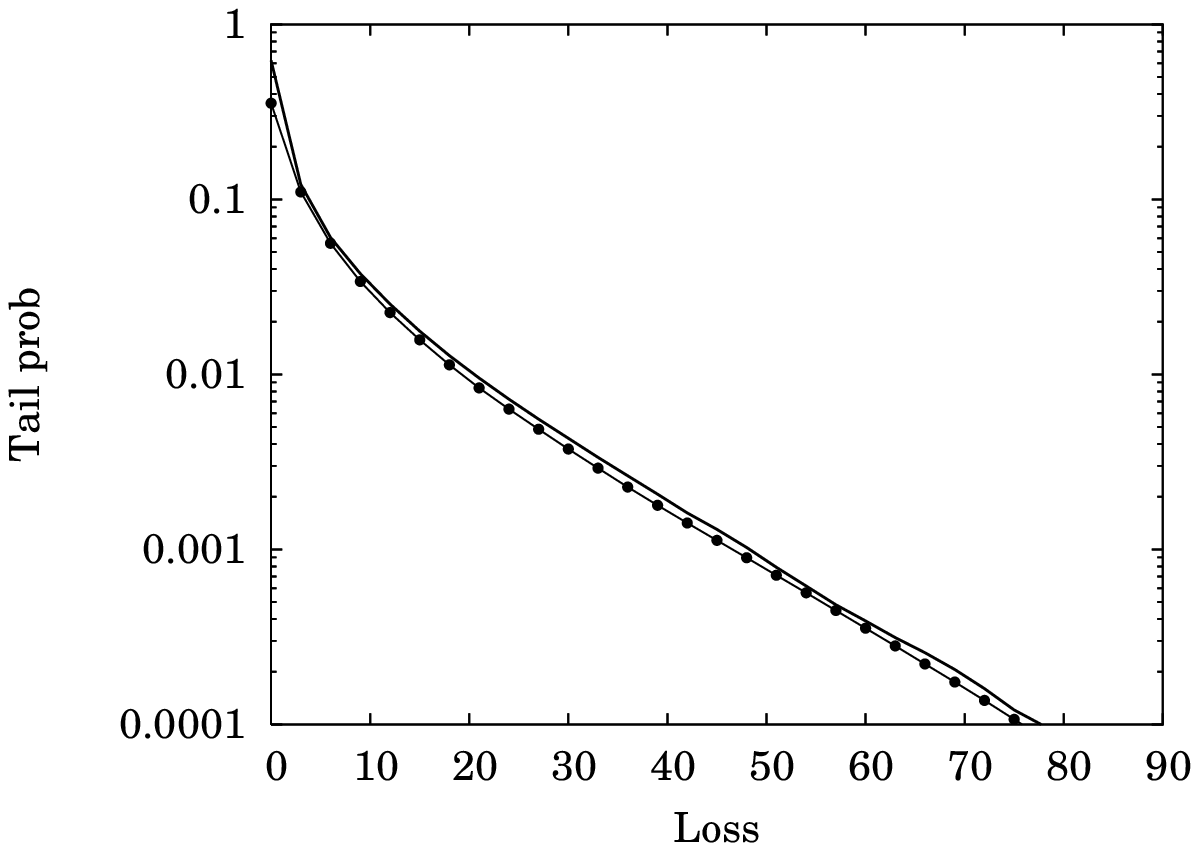}} &
\scalebox{0.6}{\includegraphics*{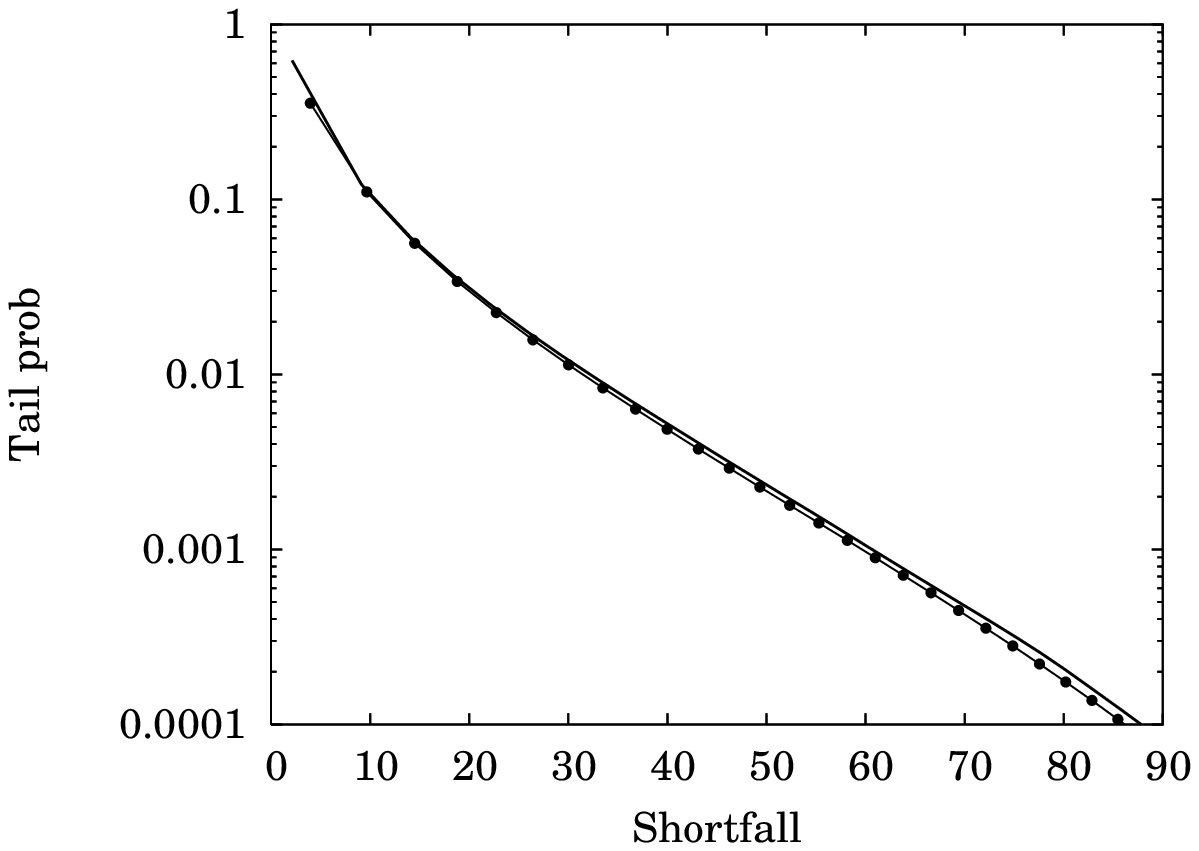}} \end{tabular}
\caption{\small As above but with $\beta=0.7$.}
\label{fig:pftestc3}
\end{figure}

\begin{figure}[h!]
\begin{tabular}{cc}\scalebox{0.6}{\includegraphics*{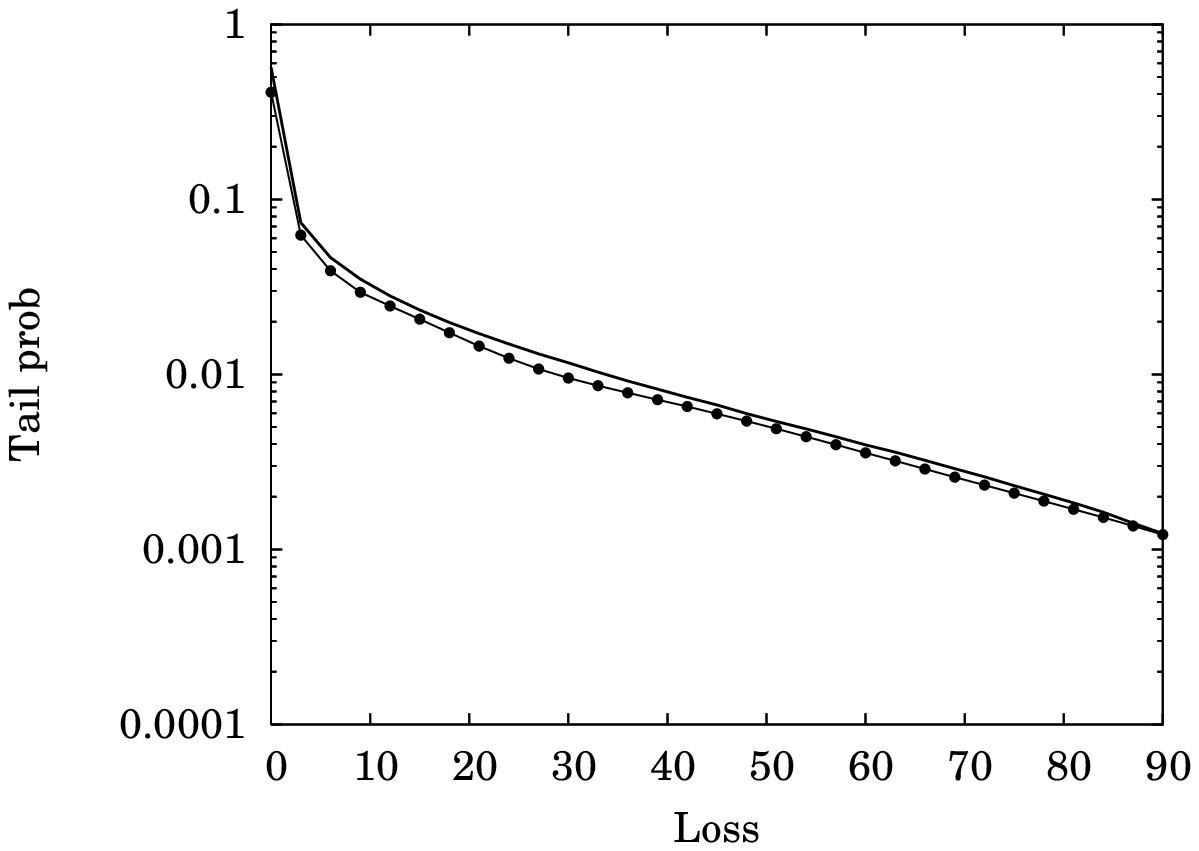}} &
\scalebox{0.6}{\includegraphics*{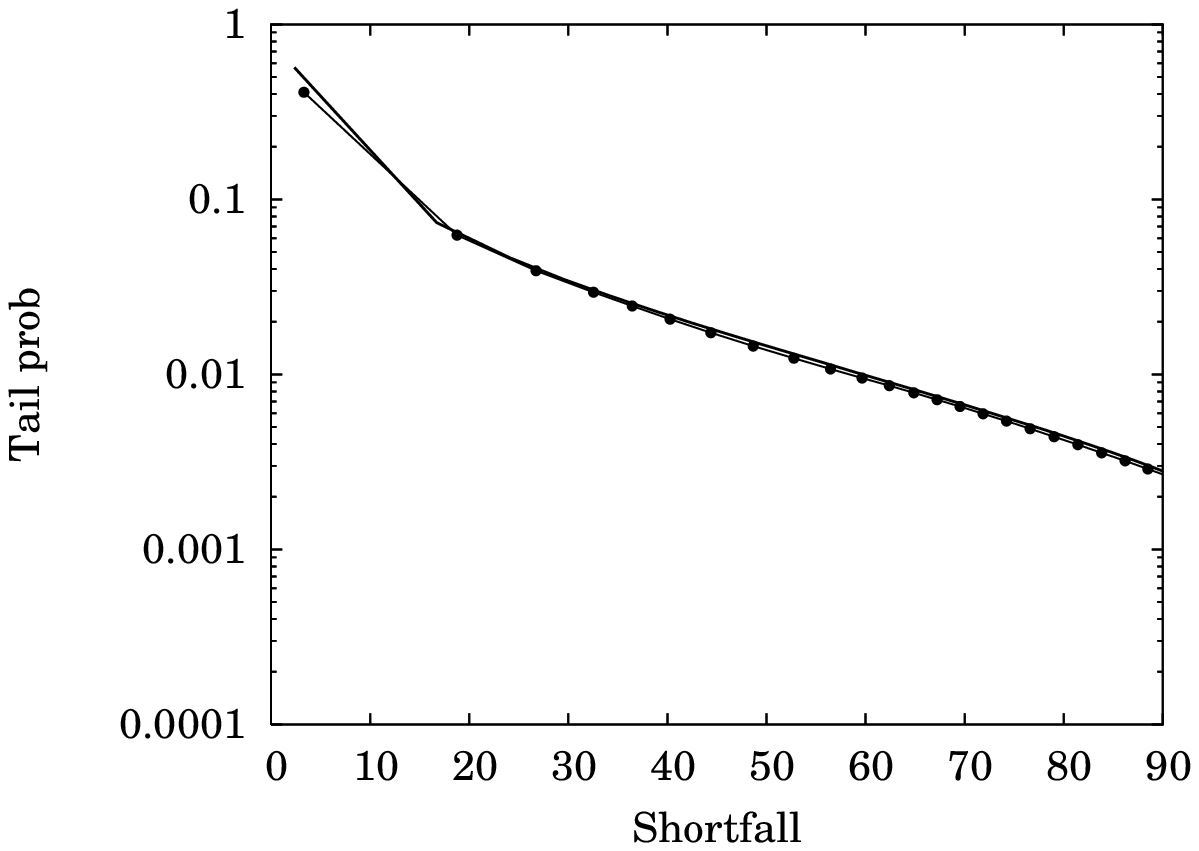}} \end{tabular}
\caption{\small As above but with $\beta=0.9$.}
\label{fig:pftestc4}
\end{figure}

As an example of correlated assets we take a default/no-default model under the one-factor Gaussian copula. (To remind: this means that the conditional default probability of an asset is $p(V)=\Phi\Big(\frac{\Phi^{-1}(\pbar)-\beta V}{\sqrt{1-\beta^2}}\Big)$,
where $V\sim \mathrm{N}(0,1)$ is the risk-factor, $\pbar$ is the expected default frequency and $\beta$ is its correlation.) The characteristics are: 50 assets, exposures net of recovery mainly 1--5 units, with one or two larger ones; $\pbar$ 0.2\%--3\%; $\beta$'s equal.
Figures~\ref{fig:pftestc1}--\ref{fig:pftestc4} show the results, comparing with Monte Carlo with 1 million simulations. Four different values of $\beta$ are considered to show the full range of correlations that might reasonably be applied in practice. 
It is not surprising that the accuracy is good, as all that is happening (in essence) is that results analogous to those displayed previously are being mixed together for different values of the risk-factor.

\subsection{The direct saddlepoint approximation}

So far our exposition of analytical approximations has been confined to sums of independent random variables, and extension to conditionally-independent variables has been trivial (condition on risk-factor, find distribution, integrate out)---which is the intuitively obvious way of doing it.
Experience shows however that the direct approach (\ref{eq:invdirect}), where the saddlepoint method is applied to the unconditional MGF, can also be quite effective. The main attraction of the direct approach is that if $C_Y(\omega)=\ex[C_{Y|V}(\omega)]$ is known in closed form, there is no subsequent integration over $V$ to be done; and the saddlepoint no longer depends on the risk-factor, which makes for simple computation. (The formulas are obvious, as one simply uses the unconditional MGF in the cases derived for independence.)
The direct approach also has the mathematical elegance of being a natural extension of mean-variance theory, which is recovered in the limit $\sp\to0$ (see \CITE{Martin01b,Martin01d}); the indirect approach does not do this, because the saddlepoint is a function of the risk-factor.
But how justifiable is the direct approach?

Suppose, for the purposes of illustration, that it is the \emph{exposures} rather than the default probabilities that depend on the risk-factor (so that correlated exposures are being modelled, as perhaps in a model of counterparty risk in OTC derivatives). Let us also assume the risk-factor to be lognormally distributed and that each counterparty's exposure is a constant multiple of that factor.  
Now focus on what happens when the factor integration is done. In the indirect approach it is the conditional tail density, probability, etc., that is being integrated over the risk-factor, which causes no problems. But in the direct approach, it is the conditional MGF that is being integrated prior to the saddlepoint approximation being done: and that integral doesn't exist, because the lognormal distribution is too fat-tailed: the integral is 
\[
M_Y(s) = \ex[e^{sY}\,|\,V] = \int_0^\infty \prod_j \big(1-p_j+p_je^{s\,\overline{a}_j\alpha}\big) \psi(\alpha) \, d\alpha
\]
where $p_j$ are the default probabilities and $\overline{a}_j$ the average exposures, and $\psi$ denotes the lognormal density.
So in this sort of situation the direct approach cannot possibly work.
Therefore the indirect approach is \emph{prima facie} more generally applicable.

On the other hand the direct method does work very well for some models:

\begin{itemize}
\item
Feuerverger \& Wong \CITE{Feuerverger00} used it to approximate the distribution of a quadratically-transformed multivariate Gaussian variable, which they used as a general tool for market risk problems.
\item
Gordy \CITE{Gordy02} used it on the CreditRisk+ model, which we have described:  the loss distribution is a Poisson distribution whose mean is stochastic and Gamma-distributed.
\item
Previously to both, Martin (\CITE{Martin98a}; see also \CITE{Jensen95}) used it to approximate a Gamma distribution integrated over a Poisson, as a prototypical insurance or reliability problem. Explicitly: events occur as a Poisson process and each time an event occurs a loss is generated. The distribution of total loss over some time period is required. In other words the distribution of $\sum_{i=1}^N X_i$ is required, where $X_i$ are i.i.d.\ $\mathrm{Gamma}(\alpha,\beta)$ and $N$ is $\mathrm{Poisson}(\theta)$. By conditioning on $N$ and integrating out, we find the MGF to be
\[
M_Y(s) = \sum_{r=0}^\infty \frac{e^{-\theta}\theta^r}{r!} (1-\beta s)^{-r\alpha} = \exp\!\left(\theta\big((1-\beta s)^{-\alpha}-1\big)\right)
\]
and, rather conveniently, the equation $K'_Y(\sp)=y$ can be solved for $\sp$ algebraically. As a check, the loss distribution can also be calculated directly as an infinite series of incomplete Gamma functions.
\end{itemize}
The first problem can be boiled down to the sum of independent Gamma-distributed variables, and in the other two, the distributions at hand are from exponential families that are known to be well-approximated by the saddlepoint method. By `exponential family' one means a distribution in which $\nu K(s)$ is a valid KGF for all real $\nu>0$. Such distributions are then infinitely divisible\footnote{Which connects them to L\'evy processes, an issue which the reader may wish to pursue separately \CITE{Sato02}.}.
It is therefore not surprising that the direct saddlepoint approximation worked well. However, there is no uniform convergence result as for a sum of independent random variables.

This is a difficult area to make general comments about. We suggest that the direct method be applied only to exponential families. This includes what is mentioned above, or for example portfolio problems in which the joint distribution of assets is multivariate Normal Inverse Gaussian or similar.

%%%%%%%%%%%%%%%%%%%%%%%%%%%%%%%%%%%%%%%%%%%%%%%%%%%%%%%%%%%%%%%%%%%%%%%%%%%%%%%%%%%%%%%%%%%%%%%%%%%%%%%%%%%%%%%%%%%%%%%%%%%%%%%%%%%%%%%%%%%%%%%%%%%%
%%%%%%%%%%%%%%%%%%%%%%%%%%%%%%%%%%%%%%%%%%%%%%%%%%%%%%%%%%%%%%%%%%%%%%%%%%%%%%%%%%%%%%%%%%%%%%%%%%%%%%%%%%%%%%%%%%%%%%%%%%%%%%%%%%%%%%%%%%%%%%%%%%%%

\section{Risk contributions}

The objective behind risk contributions is to understand how risk depends on asset allocation.
Often, the precise risk number for a portfolio is not the most important issue, as for one thing that is critically dependent on the parameters such as default probabilities or spread volatilities and distributions, default or spread correlations, and the like. It is therefore more important to use the risk measure comparatively, i.e.\ compare two portfolios or one portfolio over time. In other words, the real issue is: What makes the risk change?
This means that we wish to calculate derivatives of VaR and ESF with respect to the asset allocations in the portfolio. It turns out that these quantities provide an explicit answer to the following fundamental question: Given that I lose more than a certain amount, which assets, positions or scenarios are likely to have been responsible? We shall see that these so-called `contributions' sum to give the portfolio risk. From the point of view of the portfolio manager, this information is easily presented as a list, in decreasing order, of the biggest contributions, or, if you like, the biggest `headaches'. Once these have been identified, it is clear how to improve the efficiency of the portfolio: one identifies the riskiest positions and, if these are not generating a commensurate expected return, one hedges them or chops them down.

\begin{figure}[h!]
\centerline{\scalebox{0.6}{\includegraphics*{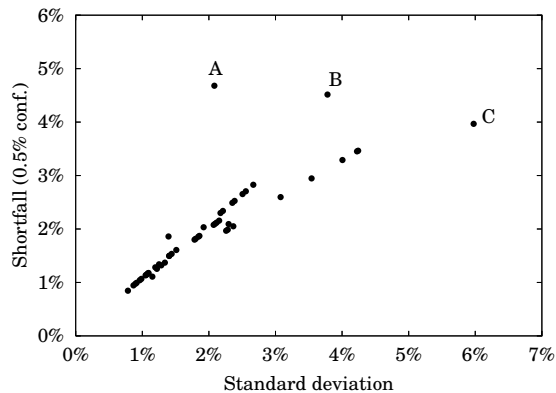}}}
\caption{\small VaR and shortfall contributions, as \% of portfolio risk, compared in a default/no-default model. `A' and `B' are tail risks (large exposures to high-grade assets) and `C' is the opposite: a smallish exposure to a very low-grade credit. In a standard deviation based optimisation, C generates most risk, with A being of little importance; in a tail-based one, A contributes most risk.}
\label{fig:cmpctrb}
\end{figure}

In the context of a credit portfolio, which instruments generate the most risk? Clearly large unhedged exposures to high-yield names will be the biggest, but there is more to it than that. For one thing, the more correlated a name is with the portfolio, the more risk it will contribute. Secondly, what is the tradeoff between credit quality and exposure? It is common sense that banks will lend more to so-called `less-risky' names than riskier ones, usually based on credit rating. But should one lend twice as much to a AA as to a BB, or twenty times? The answer to that depends on the risk measure. For tail-based measures such as VaR and ESF, poor-quality assets are not penalised as much as they are with the standard deviation measure.
This is because the worst that can happen to any credit is that it defaults, regardless of credit rating; so with VaR at 100\% confidence, the risk contribution is just the exposure (net of recovery). So a VaR- or shortfall-based optimisation will produce a portfolio with a lower proportion of so-called high-grade credits than a standard-deviation-based one, preferring to invest more in genuinely riskfree assets such as government bonds or in `rubbish' such as high-yield CDO equity that has a small downside simply because the market is already pricing in the event of almost-certain wipeout. To give a quick demonstration of this, Figure~\ref{fig:cmpctrb} compares standard deviation contribution with shortfall contribution in the tail for a default/no-default model. Tail risks and `standard deviation risks' are not equivalent.

In the context of a trading book, the considerations are different. Books are not generally conceived as portfolios, particularly on the sell-side, and positions may be turned over rapidly (potentially within a day). Given this, there is no interpretation of any position having an expected return, which would make an optimisation seemingly pointless. However, the concept of risk contribution is still important, even if risk is now to be defined as a \emph{mark-to-market} variation  at a much shorter time horizon such as a couple of weeks; default is still included, as a particularly large move in value, but for most credits will generally be too unlikely to show up on the `risk radar'. The risk contribution is to be interpreted as a way of determining positions that contribute far too much risk in relation to the profit that they might conceivably generate, such as levered positions on so-called `safe' names.

One would expect that, given (almost-) closed-form expressions for tail probability and shortfall, it should be relatively simple to obtain derivatives either by direct differentiation of the approximated tail probability and shortfall, or alternatively by performing the saddlepoint approximation on the derivatives. It turns out that the second route gives slightly neater results and that is the one that we shall pursue here.
Following option-pricing parlance, we refer to the first and second derivatives of risk measures w.r.t.\ their asset allocations as their `delta' and `gamma'.

We start by developing the VaR contribution as a conditional expectation. It turns out that there are some fundamental conceptual difficulties with VaR contribution, notably for discrete-loss models such as CreditRisk+. Paradoxically, the exact contribution is an ill-posed problem and analytical approximations are considerably more useful in practice. The shortfall contribution suffers less from these problems. For the shortfall it is known that the measure is convex, and it turns out that the saddlepoint approximation preserves that convexity. This is an important result because it shows that use of the saddlepoint-ESF will give a unique optimal portfolio in an optimisation. We will show examples along the way.

\subsection{VaR}

Our starting-point is the expression for the upper tail probability\footnote{$\mathcal R$ denotes the real axis, with $+$ indicating that the contour is indented so as to pass anticlockwise at the origin, and $-$ clockwise.}
\begin{equation}
P^+ \equiv \pr[Y\gtrdot y] = \frac{1}{2\pi\I} \invintrp C_Y(\omega) e^{-\I \omega y} \frac{d\omega}{\omega} 
\label{eq:tp_Fourier}
\end{equation}
and we now wish to perturb the VaR ($=y$) while keeping the tail probability constant. This gives
\[
0 = \deriv{P^+}{a_j} = \blobi \invintr \left(\deriv{C_Y}{a_j}-\I\omega \deriv{y}{a_j}C_Y(\omega)\right) e^{-\I \omega y} \frac{d\omega}{\omega}
\]
(the integrand is regular, so the contour can go through the origin).
As
\begin{equation}
\deriv{C_Y}{a_j} = \deriv{}{a_j} \ex[e^{\I\omega \sum_j a_jX_j}] = \I\omega \ex[X_j e^{\I\omega Y}]
\label{eq:Cderiv1}
\end{equation}
we have
\begin{equation}
\deriv{y}{a_j} = \frac{\ex[X_j\delta(Y-y)]}{\ex[\delta(Y-y)]} = \ex[X_j \cdl Y\seq y]
\label{eq:VaRdelta}
\end{equation}
which is the well-known result.
From 1-homogeneity of VaR we should have $\sum_j a_j \partial y/\partial a_j = y$, which is clear from the above equation. The VaR contribution is (\ref{eq:VaRdelta}) multiplied by $a_j$, so the VaR contributions add to the VaR.

Before pressing on with the theory let us consider some practical implications of this result. First, as the VaR contribution requires one to condition on a precise level of portfolio loss, it is quite difficult to estimate by Monte Carlo simulation. Two approaches that have been applied are kernel estimation, which uses information from simulations where the portfolio loss is close to the desired level, and importance sampling, in which the `region of interest' is preferentially sampled by changing measure. A strength of the analytical approaches is that they are free of Monte Carlo noise.

A second issue is not so well-known but fundamentally more worrying: even if it were easily computable, the exact VaR contribution might not be a very sensible construction.
The problem occurs with discrete distributions, for example in grid-based models such as CreditRisk+ \CITE{Lehrbass04}, or CDO pricing algorithms \CITE{Andersen03,Burtschell05} and in Monte Carlo simulation of any model. The following example is quite illuminating: with a default/no-default model of portfolio loss, let the exposures net of recovery be
\[
9,8,18,9,8,20,17,16,12,12.
\]
The following assertions may come as a surprise:
\begin{itemize}
\item
At portfolio loss=40, the VaR contribution of the first asset is zero (regardless of the default probabilities or correlation model). It is possible for a loss of 40 to occur (16+12+12, 8+20+12, etc.) but none of the relevant combinations includes a loss of 9.
\item
However, the VaR contribution increases if the exposure is \emph{decreased} by 1 unit (assuming that the VaR does not change thereby): 8+20+12=40. So VaR contribution is not a `sensible' function of exposure. \item
Also, the first asset does contribute if the loss level is 38 (9+9+20, etc.) or 41 (9+20+12, etc.). So VaR contribution is not a `sensible' function of attachment point either.
\item
As a consequence of the previous point, the position changes markedly if another asset is added: for example, addition of another asset will in general cause the VaR contribution of the first asset to become nonzero (e.g.\ an additional exposure of 1 gives 9+20+12+1=40).
\end{itemize}

\begin{figure}[h!]
\begin{tabular}{cc}\scalebox{0.6}{\includegraphics*{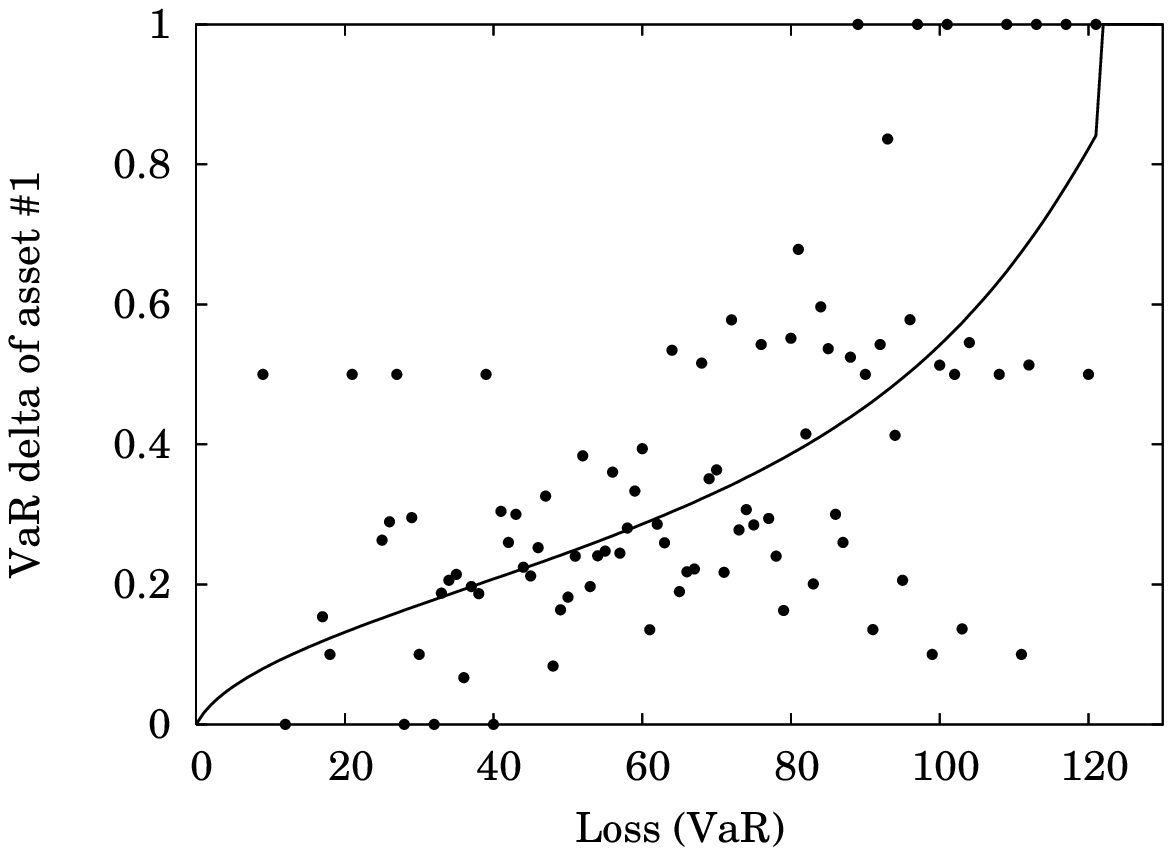}} &
\scalebox{0.6}{\includegraphics*{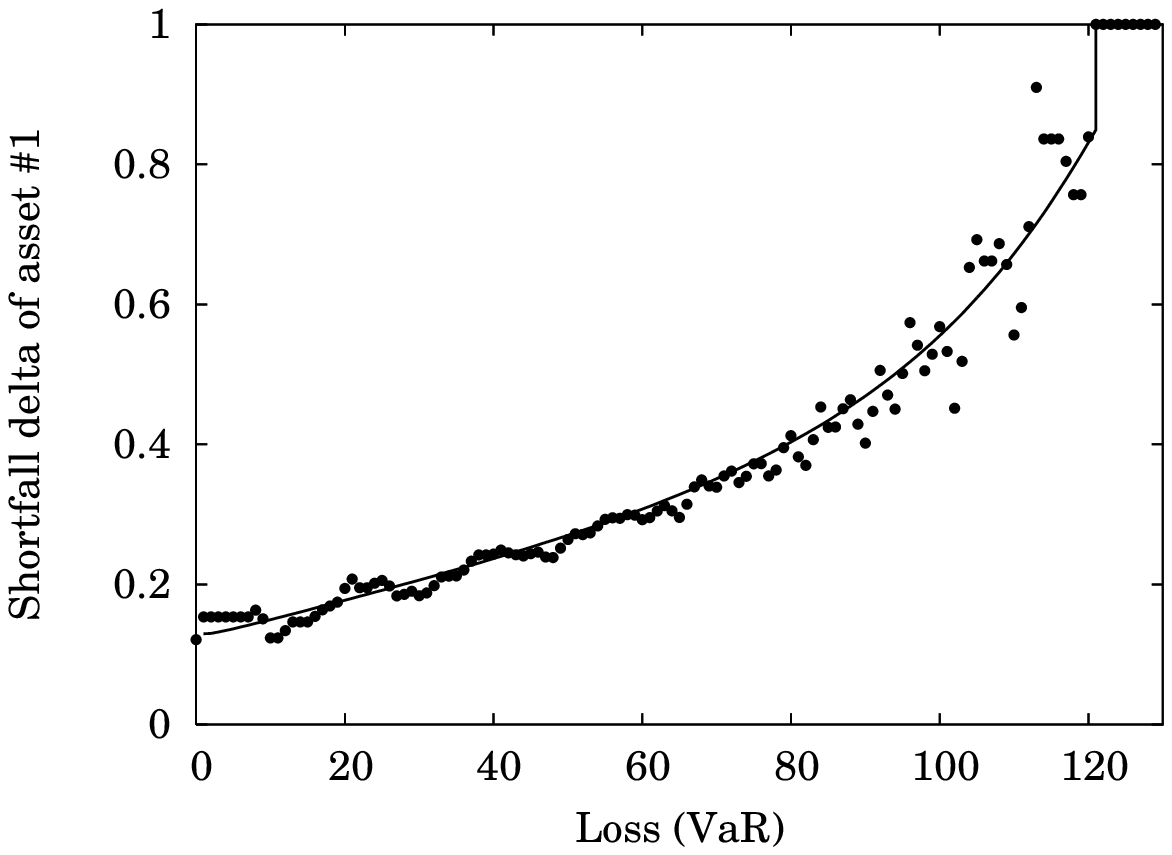}} \end{tabular}
\caption{\small VaR and shortfall contributions of one particular asset as a function of VaR. Dots show the exact result; the curve is the saddlepoint result.}
\label{fig:rc1a}
\end{figure}

\begin{figure}[h!]
\begin{tabular}{cc}\scalebox{0.6}{\includegraphics*{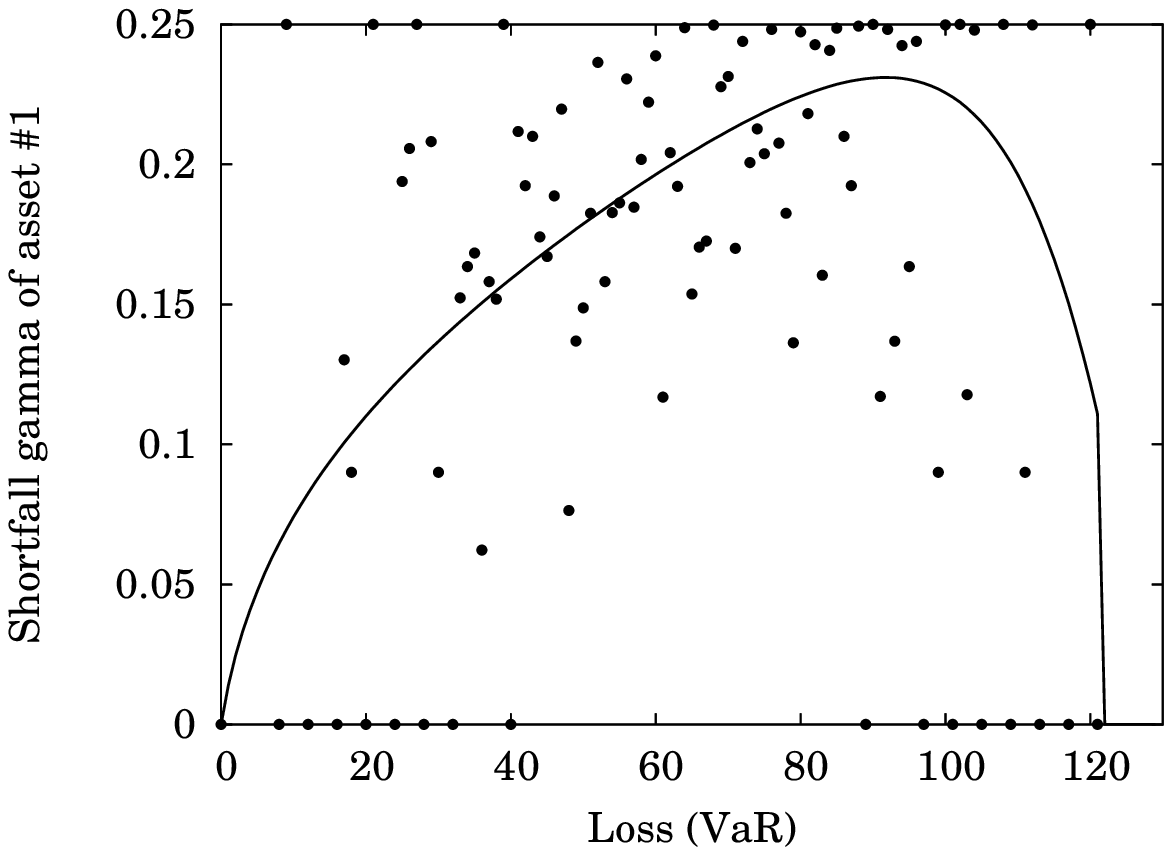}} &
\scalebox{0.6}{\includegraphics*{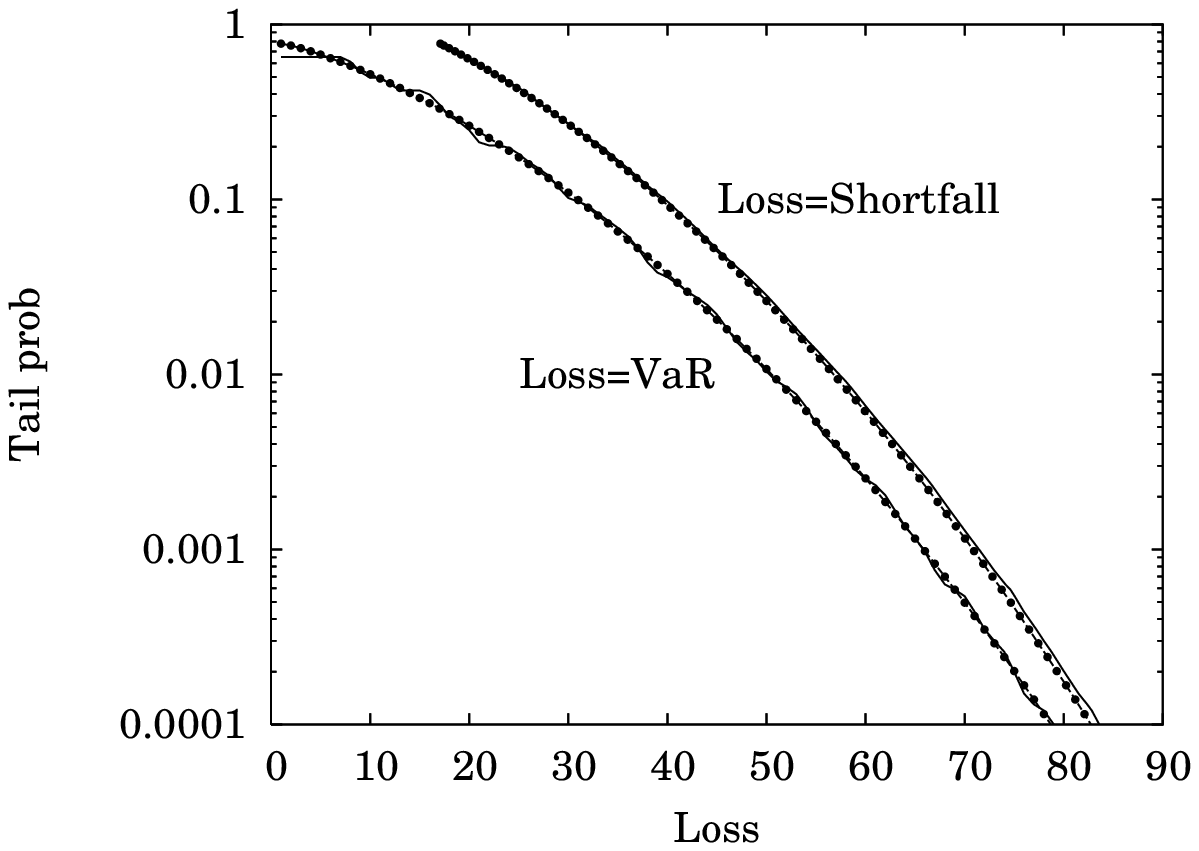}} \end{tabular}
\caption{\small (Left) Shortfall gammma of one particular asset as a function of VaR. Dots show the exact result; the curve is the saddlepoint result. (Right) 
VaR and shortfall in the portfolio. For each, the dotted line shows the approximation, and the solid one shows the exact result.}
\label{fig:rc1b}
\end{figure}

Figure~\ref{fig:rc1a} illustrates the problems for this portfolio. For simplicity we have assumed independence of the losses and given each asset a default probability of 10\%. The exact VaR deltas can be computed by direct calculation of the Fourier integral above, using the Fast Fourier Transform Algorithm. The figure shows the VaR delta for the first asset, for different loss levels. The most striking thing is that the graph is, to put it colloquially, all over the place. The delta is not even defined at some loss levels because those losses cannot occur (e.g.\ loss of 11). By way of motivation, let us point in Figure~\ref{fig:rc1a} to some more results that we shall derive presently. First, we have plotted the saddlepoint VaR delta for asset 1 as a function of loss level, and observe that the graph is nice and smooth, appearing to steer some sort of median course---and indeed, a polynomial fit to the exact VaR delta is quite close. The adjacent plot shows the shortfall deltas, which are much less `random', and again the saddlepoint approximation picks up the important feature of the graph very well. Similar remarks apply to the second derivative (`gamma') of shortfall in Figure~\ref{fig:rc1b}.
We have also plotted VaR and shortfall as functions of tail probability to compare the saddlepoint approximation with the exact answer, and find them to be very close, despite the small size (10 names) and inhomogeneity of the portfolio.

%CUT
%These problems are germane to discrete models and they do not disappear when the grid is made finer: there is no mechanism for smoothing things out. To some extent they can be obviated by averaging the contributions from `nearby' loss levels, rather as in the kernel estimation technique, but it is not clear what kernel width to use, and the closer the calculation is to the `correct' answer, the \emph{worse}-behaved it will be. 

Intriguingly, working with the VaR of a continuum model of a portfolio distribution actually works much better in practice. This leads us to the saddlepoint approximation. Incidentally Thompson and coworkers \CITE{ThompsonK03a} arrive at essentially the same equations through an application of the statistical physics concept of an ensemble, in which one looks at an average of many independent copies of the portfolio.
%The VaR contribution can easily be evaluated explicitly for any conditional independence model in which the portfolio distribution, conditional on the risk-factor, is assumed to be Normal (a reasonable idea given the CLT). 
We recast (\ref{eq:tp_Fourier}) using the MGF, writing the $V$-expectation outside the inversion integral (i.e.\ following the `indirect' route):
\begin{equation}
P^+ = \ex\left[\blobi \invintip e^{K_{Y|V}(s)-sy} \,\frac{ds}{s} \right].
\end{equation}
Differentiating w.r.t.\ the asset allocations gives
\[
0 = \deriv{P^+}{a_j} = \ex\left[\blobi \invinti \left(\deriv{K_{Y|V}}{a_j}-s\deriv{y}{a_j}\right) e^{K_{Y|V}(s)-sy}\,\frac{ds}{s} \right]
\]
(note that the integrand is now regular at the origin). Performing the saddlepoint approximation, we arrive at
\begin{equation}
\deriv{y}{a_j} \sim \frac{1}{f_Y(y)}  \ex\left[\frac{1}{\sp_V}\deriv{K_{Y|V}}{a_j}f_{Y|V}(y)\right].
\label{eq:VaRd_sp}
\end{equation}
If we define the tilted conditional expectation $\nE_V[\cdot]$ by
\begin{equation}
\nE_V[Z] = \left.\frac{\ex[Ze^{sY}\cdl V]}{\ex[e^{sY}\cdl V]}\right|_{s=\sp_V}
\label{eq:Ehat}
\end{equation}
(note, when $\sp_V=0$ we have $\nE_V\equiv \ex_V$) then we have
\[
\frac{1}{\sp_V}\deriv{K_{Y|V}}{a_j} \equiv \nE_V[X_j].
\]
%Now we have already established that
%\[
%\deriv{y}{a_j} = \ex[X_j\cdl Y\seq y] = \frac{1}{f_Y(y)} \ex\!\left[\ex[X_j \cdl Y\seq y, V]\,f_{Y|V}(y)\right]
%\]
%(the last step following from Bayes' theorem)
From (\ref{eq:VaRdelta}), (\ref{eq:VaRd_sp}) and Bayes' theorem, we can associate the conditional loss with the saddlepoint approximation,
\begin{equation}
\ex[X_j \cdl Y=y,V] \sim \nE_V[X_j].
\label{eq:cdlexpect1}
\end{equation}
Of course, it is also possible to derive this directly, i.e.\ without recourse to VaR, simply by writing down the integral representation of $\ex[X_j\delta(Y-y)]$ and performing the saddlepoint approximation.

As a specific example, in the case of a default/no-default model with independence\footnote{$p_j$, $a_j$ are as before the default probabilities and losses.} we have the contribution of the $j$th asset as
\[
\tilde{p}_j a_j, \qquad \tilde{p}_j \equiv \frac{ p_j e^{a_j \sp} }{1-p_j+p_j e^{a_j \sp}}
\]
which is its expected loss in the exponentially tilted probability measure with tilting factor $a_j\sp$. It is important to note that the default probability is thereby weighted with a loading that \emph{increases exponentially with loss}, making large loss events show up on the `risk radar' even if they have low probability. The same idea is used in importance sampling \CITE{Glasserman05}, in which tilting the probabilities to increase them allows the point of interest in the loss distribution to be preferentially sampled. Indeed, the change of measure thereby performed is that which moves the expected loss of the distribution to the desired value while keeping the distribution's entropy as high as possible (another concept that links to statistical physics). When correlation is introduced the situation is not much different, as is apparent from (\ref{eq:VaRd_sp}), for one is only averaging a similar expression over different states of the risk-factor.

It is worth noting that, for each value of the risk-factor, the saddlepoint VaR `delta' $\nE_V[X_j]$ is an increasing function of the threshold $y$ (because $K_{Y|V}(s)=\sum_j K_{X_j|V}(a_js)$ in that case and $K_{X_j|V}'$ is an increasing function). But this is a property of saddlepoint approximations, not a general result: as we already know from Figure~\ref{fig:rc1a}, where for simplicity the losses are independent, the exact VaR contribution is not necessarily an increasing function of the threshold. So the saddlepoint approximation is doing some sort of desirable `smoothing'. For positively correlated losses the saddlepoint approximation will again give monotonicity, but if they are negatively correlated then this will no longer be so: an asset negatively correlated with the portfolio will on average have a lower value when the portfolio value is high, and vice versa. (The reason for the non-monotonicity is that $f_{Y|V}(y)/f_Y(y)$ is not necessarily an increasing function of $y$; for the case of independence though, it is always unity.)

A final advantage of the saddlepoint approximation is its computational efficiency. Surprisingly perhaps, the calculation of saddlepoint risk contributions requires no more work than calculating the risk in the first place. The reason for this is that, by conditional independence,
\[
K_{Y|V}(s) = \sum_j K_{X_j|V}(a_j s)
\]
and so
\[
\frac{1}{\sp_V} \deriv{K_{Y|V}}{a_j}(\sp_V) = K'_{X_j|V}(\sp_V).
\]
Now, when $K'_{Y|V}(\sp_V)=y$ was solved for $\sp_V$, the derivatives of $K_{X_j|V}$ were calculated (the first derivative is of course essential; second and higher order  ones are used for Newton-Raphson iterative schemes). One only has to cache these derivatives (for each $j$ and $V$) during the root-searching routine and recall them when it is desired to calculate the VaR contribution. The same remark applies to both derivatives shortfall, when we derive them.
This compares favourably with other methods in which differentiation must be carried out for each asset in the portfolio \CITE{Takano08}.

To round this section off we show the special case where $\sp_V\to0$, which generates the CLT. If the portfolio loss distribution conditional on $V$ is Normal with mean $\mu_{Y|V}$ and variance $\sigma^2_{Y|V})$, then the conditional density and distribution function are, respectively,
\[
f_{Y|V} = \ex[\sigma^{-1}_{Y|V}\phi(z_V)], \qquad
\pr[Y<y] = \ex[\Phi(z_V)],
\]
with
\[
z_V \equiv \frac{y-\mu_{Y|V}}{\sigma_{Y|V}}.
\]
(note that this \emph{is} actually the same $z$ as we used in conjunction with the Lugannani-Rice formula earlier). It is then readily established, and also verifiable from first principles, that
\begin{equation}
\deriv{y}{a_j} =
\frac{1}{f_Y(y)} \ex\left[
\left(\deriv{\mu_{Y|V}}{a_j} + z_V\deriv{\sigma_{Y|V}}{a_j}\right)\frac{1}{\sigma_{Y|V}} \phi(z_V)
\right].
\label{eq:VaRd_CLT}
\end{equation}

%We are now ready to tackle the expected shortfall measure.

\subsection{Shortfall}

\subsubsection*{First derivative}

The starting-point is the Fourier representation of $\sh^\pm[Y]$:
\begin{equation}
\ex[Y\cdl Y\!\gtrless\! y] = y \mp \frac{1}{2\pi P^\pm} \invintrpm C_Y(\omega) e^{-\I\omega y} \, \frac{d\omega}{\omega^2}
\label{eq:esf_Fourier}
\end{equation}
From this we obtain derivative information via the obvious route of differentiating w.r.t.\ $a_j$ at constant tail probability. Note that because the tail probability is being held fixed, $y$ will depend on the $(a_j)$; however, we know all about that from the VaR derivations, and conveniently some cancellation occurs en route. This gives
\[
\deriv{\sh^\pm[Y]}{a_j} = \mp\frac{1}{2\pi P^\pm} \invintrpm \deriv{C_Y}{a_j} e^{-\I\omega y}\, \frac{d\omega}{\omega^2}.
\]
Using (\ref{eq:Cderiv1}) we obtain 
\begin{eqnarray}
\deriv{\sh^\pm[Y]}{a_j} = \frac{\pm1}{2\pi\I P^\pm} \invintrpm \ex[X_je^{\I\omega Y}] e^{-\I\omega y} \, \frac{d\omega}{\omega} \nonumber\\
= \frac{1}{P^\pm} \ex\big[X_j\indic{Y \gtrless y}\big]
= \ex\big[X_j\cdl Y\!\gtrless\!y].
\label{eq:ESFdelta}
\end{eqnarray}
It is clear that the homogeneity result is obeyed: $\sum_j a_j\partial\sh^\pm[Y]/\partial a_j=\sh^\pm[Y]$.

We said when discussing the problems of VaR contribution for discrete distributions that ESF is smoother than VaR, and in fact it is one order of differentiability smoother. This means that the first derivative of ESF is a continuous function of asset allocation, but, as we shall see later, the second derivative of ESF is as `bad' as first derivative of VaR (so it is discontinuous). So we still want to work everything through for the analytical approximations, which is what we do next.

 The saddlepoint analogue of the equations is
\begin{equation}
\deriv{\sh^\pm[Y]}{a_j} =  \frac{\pm 1}{2\pi\I P^\pm} \ex\left[\invintipm\deriv{K_{Y|V}}{a_j} e^{K_{Y|V}(s)-sy}\, \frac{ds}{s^2}  \right]
\end{equation}
(and the integrand now has only a \emph{simple} pole at the origin).
Following \CITE{Martin06b} we split out a singular part that can be integrated exactly and leave a regular part to be saddlepoint-approximated. The result then follows\footnote{$\mu_{X_j|V}$ denotes the conditional expectation of the $j$th asset.}:
\begin{eqnarray}
\deriv{\sh^\pm[Y]}{a_j} &=&  \frac{1}{2\pi\I P^\pm} \ex\left[\pm\mu_{X_j|V} \invintipm e^{K_{Y|V}(s)-sy}\, \frac{ds}{s}
 \right. \nonumber \\
 &&\pm \left. \invintipm\frac{1}{s}\left(\frac{1}{s} \deriv{K_{Y|V}}{a_j} -\mu_{X_j|V}\right) e^{K_{Y|V}(s)-sy}\, ds \right] \nonumber \\
&=&
\frac{1}{P^\pm} \ex\!\left[\mu_{X_j|V}\pr[Y\gtrless y \cdl V] \pm \frac{1}{\sp_V} \left(\frac{1}{\sp_V}\deriv{K_{Y|V}}{a_j} - \mu_{X_j|V}\right) f_{Y|V}(y)\right] \nonumber\\
\mbox{} 
\label{eq:ESFd_sp}
\end{eqnarray}

To obtain the conditional-Normal result, we can either let $\sp_V\to 0$ or derive from first principles. As
\begin{equation}
\sh^\pm[Y] = \frac{1}{P^\pm} \ex\big[\mu_{Y|V}\Phi(\mp z_V) \pm \sigma_{Y|V}\phi(z_V)\big]
\label{eq:ESF_CLT}
\end{equation}
we have\footnote{It may appear that some of the terms have gone missing, on account of $z_V$ depending on the asset allocations; however, these terms cancel. A nice thing about working with ESF is that much of the algebra does come out cleanly.}
\begin{equation}
\deriv{\sh^\pm[Y]}{a_j} = \frac{1}{P^\pm} \ex\left[\deriv{\mu_{Y|V}}{a_j}\Phi(\mp z_V) \pm \deriv{\sigma_{Y|V}}{a_j}\phi(z_V)\right].
\label{eq:ESFd_CLT}
\end{equation}
As we pointed out earlier in the context of the shortfall at portfolio level (and in  \CITE{Martin06b,Martin07a}), shortfall expressions naturally fall into two parts, the first being systematic risk and the second unsystematic. We can see that the same holds true for the deltas,   in (\ref{eq:ESFd_CLT}) and (\ref{eq:ESFd_sp}) . 
The systematic part can be written 
\[
\ex\big[\ex_V[X_j]\cdl Y\!\gtrless\! y]\big];
\]
expressed in ungainly fashion, it is the expectation of the conditional-on-the-risk-factor-expectation of an asset, conditionally on the portfolio losing more than the VaR.
The discussion surrounding Figure~\ref{fig:rc2b} later will show this in practice: correlated assets or events contribute strongly to the first part, which roughly speaking is proportional to the asset's `beta', and uncorrelated ones mainly to the second.

Note also that the contribution to unsystematic risk \emph{must be positive}, for both approximations, and for the exact result too, because conditionally on $V$ the assets are independent and so any increase in allocation to an asset must cause the conditional variance (= unsystematic risk) to increase. This is exactly what we would like to conclude (see \CITE{Martin07a} for a fuller discussion). VaR does not have this property, which means that one could in principle be in the awkward position of increasing the exposure to an uncorrelated asset, keeping the others fixed, and watching the VaR \emph{decrease}.

\subsubsection*{Second derivative}

One of the main things that we shall prove here is that the saddlepoint approximation preserves the convexity of the shortfall. The convexity arises in the first place because the Hessian matrix (second derivative) of the shortfall can be written as a conditional covariance, so we show that first.

Differentiating (\ref{eq:esf_Fourier}) again gives
\[
\dderiv{\sh^\pm[Y]}{a_j}{a_k} =  \frac{\mp 1}{2\pi P^\pm} \left[\invintrpm\dderiv{C_Y}{a_j}{a_k} e^{-\I\omega y}\,\frac{d\omega}{\omega^2} + \deriv{y}{a_k}\invintrpm \deriv{C_Y}{a_j} e^{-\I\omega y} \frac{d\omega}{\I\omega} \right]
\]
and the second integral is already known to us because we dealt with it when doing the VaR delta.
As
\[
\dderiv{C_Y}{a_j}{a_k} = -\omega^2 \ex[X_jX_k e^{\I\omega Y}],
\]
we can tidy things up to arrive at
\begin{eqnarray}
\dderiv{\sh^\pm[Y]}{a_j}{a_k}  &=&  \frac{\pm f_Y(y)}{P^\pm} \big(\ex[X_jX_k\cdl Y\seq y] - \ex[X_j\cdl Y\seq y]\ex[X_k\cdl Y\seq y]\big) \nonumber\\
&=& \frac{\pm f_Y(y)}{P^\pm} \V[X_j,X_k\cdl Y\seq y]
\label{eq:ESFdd}
\end{eqnarray}
where $\V[]$ denotes covariance.
Hence the second derivative of ESF is (up to a factor) the conditional covariance of the pair of assets in question, conditionally on the portfolio value being \emph{equal to} the VaR. As any covariance matrix is positive semidefinite the ESF is a convex function of the asset allocations. It is also clear that
\begin{equation}
\sum_{j,k} a_ja_k \dderiv{\sh^\pm[Y]}{a_j}{a_k} = 0,
\label{eq:ESFhg}
\end{equation}
as the LHS is, up to a factor,
\[
\sum_{j,k} a_ja_k  \V[X_j,X_k\cdl Y\seq y] = \V[Y,Y\cdl Y\seq y] = 0.
\]
This is as expected: if the risk measure is 1-homogeneous then scaling all the asset allocations by some factor causes the risk to increase linearly (so the second derivative iz zero) and so the vector of asset allocations must be a null eigenvector of the Hessian matrix. Incidentally, because the conditioning is \emph{on} a particular portfolio value or loss, the estimation of second derivative of ESF is as troublesome as that of the first derivative of VaR for discrete portfolio models or Monte Carlo.

 We now turn to the saddlepoint approximation, which as we shall see contains a trap for the unwary. The second derivative of ESF has (after using (\ref{eq:VaRd_sp}) to tidy it up) the following integral representation:
\begin{eqnarray}
\dderiv{\sh^\pm[Y]}{a_j}{a_k} &=& \frac{\pm 1}{P^\pm} \left\{ \ex\left[\blobi\invintipm\left(
\dderiv{K_{Y|V}}{a_j}{a_k} + \deriv{K_{Y|V}}{a_j}\deriv{K_{Y|V}}{a_k} 
\right)e^{K_{Y|V}(s)-sy}\, \frac{ds}{s^2}  \right]\right. \nonumber\\
&&
\left. - \deriv{y}{a_j}\deriv{y}{a_k} f_Y(y) \right\}
\label{eq:ESFdd_ci}
\end{eqnarray}
As the integrand is regular at the origin, it looks obvious to approximate the integral as
\[
\left(\frac{1}{s^2}\dderiv{K_{Y|V}}{a_j}{a_k} +  \frac{1}{s}\deriv{K_{Y|V}}{a_j}\cdot\frac{1}{s}\deriv{K_{Y|V}}{a_k}\right)_{s=\sp_V}f_{Y|V}(y). \qquad (??)
\]
This expression is incorrect and if it is used, one ends up violating (\ref{eq:ESFhg}) and producing expressions for the conditional asset covariances that do not make sense. [From now on, we abbreviate $K_{Y|V}(s)$ to $K$.] 

The problem is that we have been careless about the size of the terms neglected in the asymptotic expansion and ended up by omitting a term that is of the \emph{same} order as shown above. In recalling that the saddlepoint expansion is asymptotic as $K\to\infty$, the error becomes clear: $(\partial K/\partial a_j)(\partial K/\partial a_k)$ is order $K^2$, and therefore needs a higher-order treatment at the saddlepoint to make the neglected terms of consistent order. Suffice it to say that the correct expression for the integral is\footnote{Actually a term is still missing from this as there are two terms of the form $(\partial/\partial s)^2(s^{-1}\partial K/\partial a_j) \times s^{-1}\partial K/\partial a_k$. However, omission of these does not violate (\ref{eq:ESFhg}), and they also vanish for the Normal case.}
\[
\left\{
\frac{1}{s^2}\dderiv{K}{a_j}{a_k} +  \frac{1}{s}\deriv{K}{a_j}\cdot\frac{1}{s}\deriv{K}{a_k}
- \frac{1}{K''(s)} \left(\deriv{}{s}\frac{1}{s}\deriv{K}{a_j}\right)\left(\deriv{}{s}\frac{1}{s}\deriv{K}{a_k}\right)
\right\}_{s=\sp_V} f_{Y|V}(y).
\]

% end notthis %%%%%%%%%%%%%%%%%%%%%%%%%%%%%%%%%%%%%%%%%%%%%%%%%%%%%%%%%%%%%%%%%%%%%%%

\noindent
On collecting terms we arrive at an expression that naturally divides into two parts:
\[
\dderiv{\sh^\pm[Y]}{a_j}{a_k} \sim \frac{\pm1}{P^\pm} \ex\!\left[\big(H^S_{jk}+H^U_{jk}\big)f_{Y|V}(y)\right],
\]
or equivalently, by (\ref{eq:ESFdd}),
\[
\V[X_j,X_k\cdl Y\seq y] \sim \frac{1}{f_Y(y)} \ex\!\left[\big(H^S_{jk}+H^U_{jk}\big)f_{Y|V}(y)\right],
\]
with
\begin{eqnarray}
H^S_{jk} &=& \left(\frac{1}{\sp_V}\deriv{K}{a_j}-\deriv{y}{a_j}\right)\left(\frac{1}{\sp_V}\deriv{K}{a_k}-\deriv{y}{a_k}\right)  \label{eq:ESFdd_sp} \\
H^U_{jk} &=& \frac{1}{\sp_V^2}\dderiv{K}{a_j}{a_k} - \frac{1}{K''(\sp_V)}\left(\deriv{}{s}\frac{1}{s}\deriv{K}{a_j}\right)_{s=\sp_V}\!\!\left(\deriv{}{s}\frac{1}{s}\deriv{K}{a_k}\right)_{s=\sp_V}\nonumber
\end{eqnarray}
It is clear that $H^S$ is a positive semidefinite matrix (because it is of the form $v_jv_k$); also $\sum_{j,k}a_ja_k H^S_{jk}=0$ by (\ref{eq:VaRd_sp}).

By homogeneity properties of $K$ (it depends on $a_j$ and $s$ only through their product),
\[
\Dderiv{K}{s} =\sum_{j} a_j\deriv{}{s} \frac{1}{s} \deriv{K}{a_j} 
=\sum_{j,k} a_ja_k \frac{1}{s^2} \dderiv{K}{a_j}{a_k}
\]
so $\sum_{j,k}a_ja_k H^U_{jk}=0$ too, and (\ref{eq:ESFhg}) is satisfied.
The only thing to settle now is whether $H^U$ is positive semidefinite, and this is basically the Cauchy-Schwarz inequality. Indeed, defining the quadratic
\[
q(t) = \nE\left[\left(\big(Y-\nE[Y]\big) + t\,u\cdot(X-\nE[X])\right)^2\right]
\] 
(we drop the $_V$ suffix for convenience), where $X$ is the vector of asset values and $u$ is some arbitrary vector, we find the coefficients of $1,t,t^2$ in $q(t)$ to be:
\[
\begin{array}{rlcl}
1: & \nE\Big[\big(Y-\nE[Y]\big)^2\Big] &=& \displaystyle\Dderiv{K}{s} \\
t: & 2\nE\Big[u\cdot\big(X-\nE[X]\big)\big(Y-\nE[Y]\big)\Big] &=& \displaystyle 2\sum_j u_j \deriv{}{s}\frac{1}{s}\deriv{K}{a_j}\\
t^2: & \nE\Big[\Big(u\cdot\big(X-\nE[X]\big)\Big)^2\Big] &=& \displaystyle \sum_{j,k} \frac{u_ju_k}{s^2}\dderiv{K}{a_j}{a_k}
\end{array}
\]
As $q$ can never be negative its discriminant must be $\le0$, so
\[
\Dderiv{K}{s} \sum_{j,k} \frac{u_ju_k}{s^2}\dderiv{K}{a_j}{a_k} \ge
\left[\sum_j u_j \deriv{}{s}\frac{1}{s}\deriv{K}{a_j}\right]^2
\]
which amounts to $\sum_{j,k}u_ju_kH^U_{jk}\ge0$, as required. We conclude that the saddlepoint approximation has preserved the convexity.

In the conditional-Normal framework, we have a rather simpler expression. In the same notation as before:
\begin{equation}
\dderiv{\sh^\pm[Y]}{a_j}{a_k} = \frac{\pm1}{P^\pm} \ex\!\left[
\left(\sigma^2_{Y|V}\deriv{z_V}{a_j}\deriv{z_V}{a_k} + \sigma_{Y|V}\dderiv{\sigma_{Y|V}}{a_j}{a_k} \right) \sigma^{-1}_{Y|V}\phi(z_V) \right]
\label{eq:ESFdd_CLT}
\end{equation}
Predictably, we can identify the first term as a convexity contribution to systematic risk and the second to the unsystematic risk.
As we can expect from the saddlepoint derivation, \emph{both} terms are positive semidefinite matrices: the first is of the form $v_jv_k$ with $v_j=\sigma_{Y|V}\partial z_V/\partial a_j$, while the second is the Hessian of the standard deviation (a convex risk measure). As the mixture weights $\sigma^{-1}_{Y|V}\phi(z_V)$ are nonnegative---they are the $V$-conditional density of $Y$---the LHS must also be semidefinite.

\subsubsection*{Conditional covariance}

There are some interesting issues surrounding the covariance of two assets conditional on the portfolio loss being some level which, as we have seen, is the second derivative of shortfall.
We earlier found the approximation to the expectation of an asset conditionally on the portfolio value (again we drop the $_V$), as a tilted expectation:
\[
\ex[X_j\cdl Y=y] \sim \nE[X_j], \qquad
\nE[Z] = \left.\frac{\ex[Ze^{sY}]}{\ex[e^{sY}]}\right|_{s=\sp}.
\]
It is superficially attractive, but incorrect, to extend this to higher-order expressions, thereby surmising
\[
\ex[X_jX_k\cdl Y=y] \stackrel{??}{\sim} \nE[X_jX_k]
\]
and (defining the tilted variance in the obvious way, i.e.\ as the tilted expectation of the product minus the product of the tilted expectations)
\[
\V[X_j,X_k\cdl Y=y] \stackrel{??}{\sim} \nV[X_j,X_k].
\]
The problem with this is that if one multiplies by $a_ja_k$ and sums over both suffices one ends up with
\[
\V[Y,Y\cdl Y=y] \stackrel{??}{\sim} \nV[Y,Y].
\]
The LHS must be zero, of course (it is the variance of something that is not being allowed to vary), but the RHS is $\partial^2K_Y/\partial s^2\ne0$. In other words, the homogeneity relation is wrong.

The reason for the error is that terms have gone missing in the derivation, and those terms are the ones alluded to in the saddlepoint derivation above\footnote{The error occurs in \CITE{ThompsonK03b}, where higher-order generalisations were also stated without proof.  Correcting these seems to require a fair amount of reworking of their theory.}. The quadratic (covariance) expression should read
\[
\V[X_j,X_k\cdl Y=y] \sim \nV[X_j,X_k]-\frac{\nV[X_j,Y]\nV[X_k,Y]}{\nV[Y,Y]}.
\]
In effect, the extra term corrects the homogeneity relation by subtracting the unwanted convexity along the line of the asset allocations, exactly as in the following method of killing eigenvalues in a symmetric matrix $\Omega$:
\[
\Omega^* = \Omega - \frac{\Omega a \otimes\Omega a}{a'\Omega a}
\]
If $\Omega$ is positive semidefinite then so is $\Omega^*$ (by Cauchy-Schwarz); and as $\Omega^*a=0$, $\Omega^*$ has one more zero eigenvalue than $\Omega$.

\begin{figure}[h!]
\centerline{\scalebox{0.6}{\includegraphics*{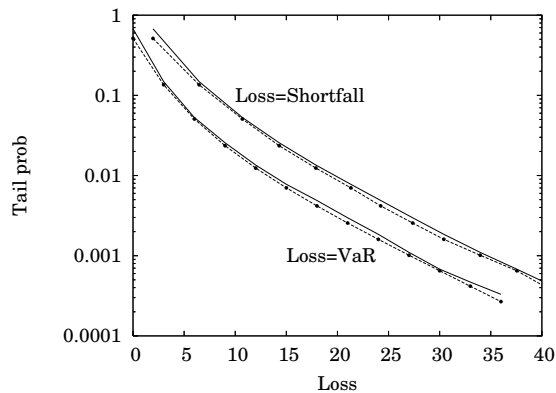}}}
\caption{\small VaR and shortfall in 50-asset portfolio. For each, the dotted line shows the approximation, and the solid one shows the result obtained by Monte Carlo.}
\label{fig:rc2a}
\end{figure}

\begin{figure}[h!]
\centerline{\scalebox{0.6}{\includegraphics*{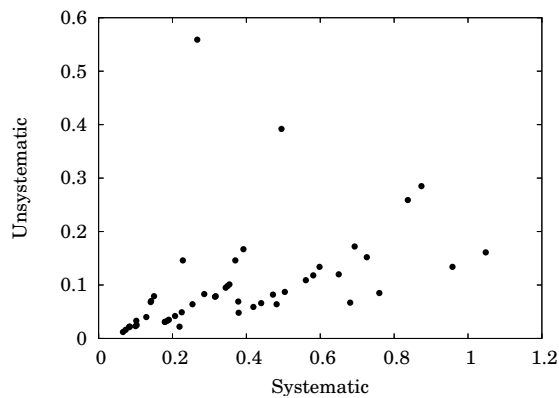}}}
\caption{\small For all the assets in the portfolio, this plot shows their shortfall contribution divided into systematic and unsystematic parts. Correlated assets sit on the bottom right, large single-name exposures on the top left. Assets in the bottom left contribute little risk.}
\label{fig:rc2b}
\end{figure}

\begin{figure}[h!]
\begin{tabular}{cc}\scalebox{0.6}{\includegraphics*{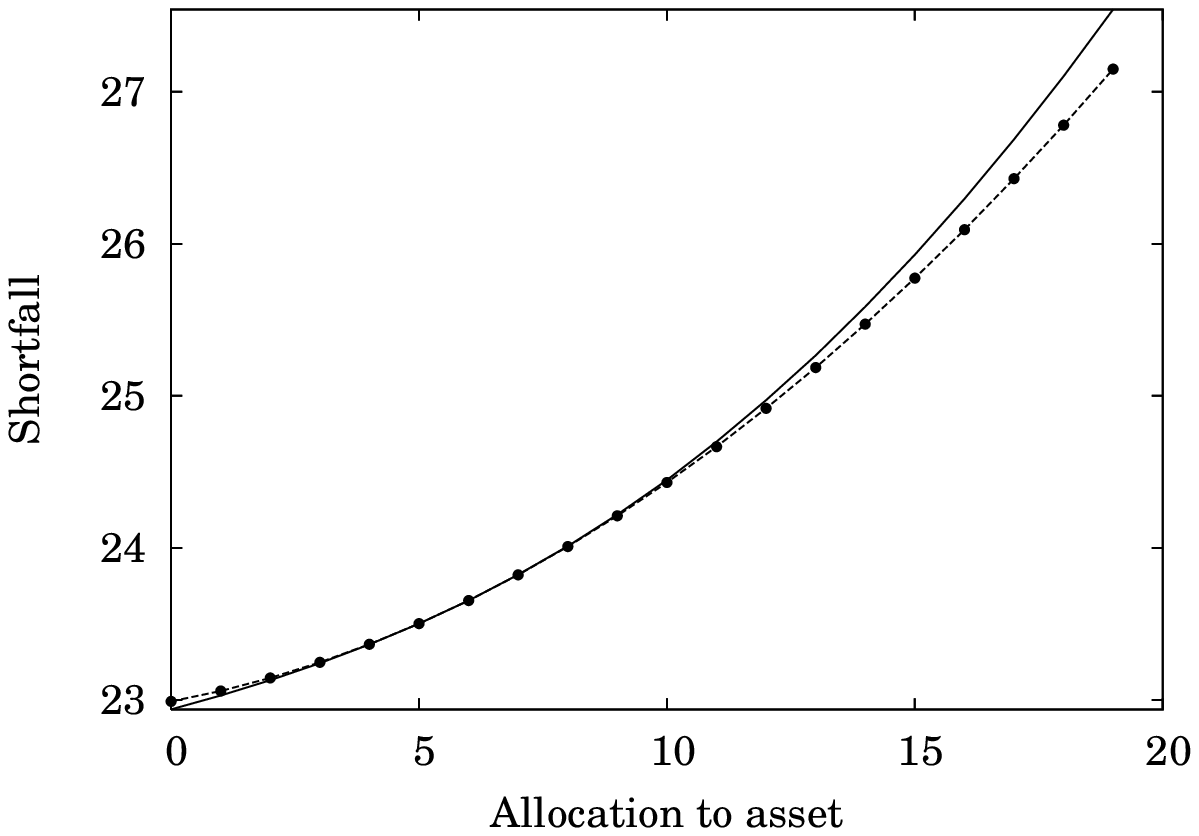}} &
\scalebox{0.6}{\includegraphics*{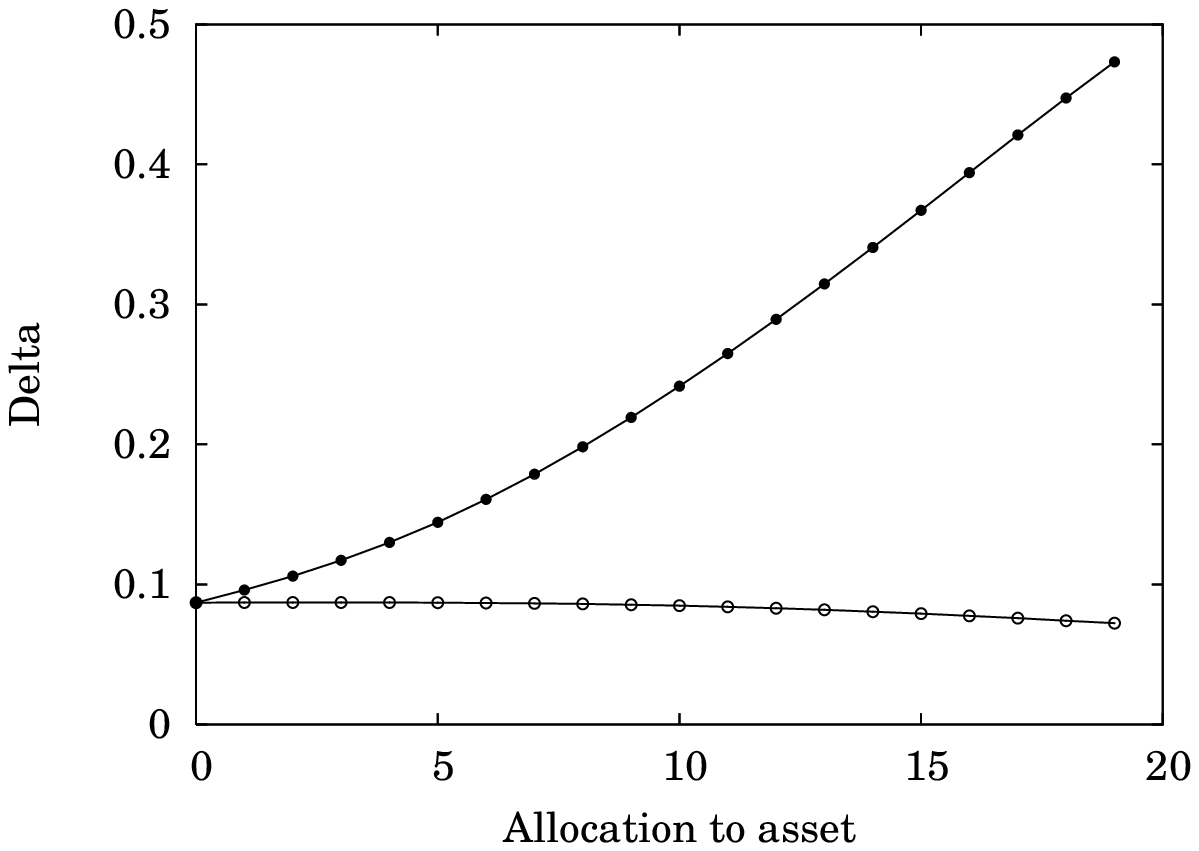}} \end{tabular}
\caption{\small [Left] Shortfall (solid line) vs quadratic approximation (dotted), as asset allocation is varied. The approximation requires only the shortfall, delta and gamma in the `base case', whereas the exact computation requires a full calculation at each point.
[Right] Again for varying that particular asset allocation, this shows the delta of the asset and the systematic delta. The systematic delta remains roughly constant so the inference is that more unsystematic risk is being added to the portfolio.}
\label{fig:rc2c}
\end{figure}

\subsection{Examples}

We conclude with some examples of the theory in action.

We revisit the example of Figures~\ref{fig:pftestc1}--\ref{fig:pftestc4} of a default/no-default model under the one-factor Gaussian copula, where now we mix up the $\beta$'s, so that some assets have $\beta=0.3$, some $0.5$ and some $0.7$. Figure~\ref{fig:rc2b} shows the systematic and unsystematic ESF contributions (\ref{eq:ESFd_sp}) of all the assets, in a scatter graph. The asset at the far top-left of the graph (coordinates $(0.27,0.56)$) is the largest exposure and is to a name with low default probability, i.e.\ a tail risk; those on the bottom right have higher correlation and higher default probability, though generally lower exposure. The interpretation of Figure~\ref{fig:rc1b} for the purposes of portfolio management is that the assets in the top left need single-name hedges and those in the bottom-right can largely be dealt with using baskets or bespoke index protection.

Next we consider what happens when one particular exposure is varied. Figure~\ref{fig:rc2c} shows the performance surface (risk vs allocation), and also shows a quadratic approximation based on the first and second derivatives estimated using (\ref{eq:ESFd_sp},\ref{eq:ESFdd_sp}). The importance of this is that the derivatives can be easily calculated along with the risk in one go, after which the quadratic approximant is easily drawn, whereas tracing the exact line requires the portfolio to be re-analysed for each desired asset allocation.
%(Arguably there are better choices than a quadratic, on the grounds that for large allocations the risk should be asymptotically linear, so one could experiment with other forms such as $\sqrt{\beta_0+\beta_1x+\half\beta_2x^2}$ and match the function and its first two derivatives to obtain the $\beta$'s. For the standard deviation risk measure, this would be exact.)

\section{Conclusions}

We have given a reasonably complete overview of the theory of risk aggregation and disaggregation of distributions that are not necessarily Normal. The results from Fourier transformation are completely general; the saddlepoint ones assume the existence of the moment-generating function, and are thereby suited to semi-heavy-tailed distributions. This assumption is sufficiently restrictive to allow useful and interesting results to be derived, particularly in the subject of risk contributions; and it is sufficiently general to allow application to real-world problems.
The  subject  has developed over the last ten years and is now gaining general acceptance. In general we prefer to operate with shortfall rather than VaR and routinely use the formulae here for the risk contributions and the like.

It is worth emphasising however that, no matter how clever the techniques of calculation, the results are only as good as the model assumptions used to provide the input.
In the `credit crunch' these have generally been found to be woefully inadequate. However, the  methods described here to perform fast aggregation and disaggregation of risk are an important part of the backbone of credit risk management systems and are to be recommended.

\subsection*{Summary of results}

To help the reader navigate the formulas, the following summarises where the results can be found. First, the basic results for the risk measures and related quantities:

\begin{center}\begin{tabular}{ll}
\hline 
& Eqn \\ 
\hline 
Density &  (\ref{eq:spdensKY}/\ref{eq:spdens_indir}) \\
Tail Probability & (\ref{eq:BN}/\ref{eq:BN_indir}) \\
Shortfall & (\ref{eq:ESF_sp}/\ref{eq:ESF_sp_indir}) \\
\hline
\end{tabular}\end{center}

\noindent Secondly, for the derivatives: 

\begin{center}\begin{tabular}{l|ccc|ccc}
\hline 
& & 1st & & & 2nd \\
& Exact & CLT & SPt & Exact & CLT &  SPt \\
\hline
VaR & (\ref{eq:VaRdelta}) & (\ref{eq:VaRd_CLT}) & (\ref{eq:VaRd_sp})  & - & - & - \\
Shortfall & (\ref{eq:ESFdelta}) & (\ref{eq:ESFd_CLT}) & (\ref{eq:ESFd_sp}) & (\ref{eq:ESFdd}) & (\ref{eq:ESFdd_CLT}) & (\ref{eq:ESFdd_sp}) \\
\hline
\end{tabular}\end{center}

\subsection*{Acknowledgements}

I have benefited greatly from interacting with other practitioners in the last ten years. Two deserve a special mention: Tom Wilde for his perceptive observations and depth of understanding in credit portfolio modelling, and Fer Koch for his particular talent in building analytics and systems and understanding how best to develop and apply the results in the rigours of a live trading environment.

\bibliographystyle{plain}
\bibliography{../phd}

\begin{thebibliography}{10}

\bibitem{Tasche02a}
C.~Acerbi and D.~Tasche.
\newblock On the coherence of {E}xpected {S}hortfall.
\newblock {\em J. Banking and Finance}, 26(7):1487--1503, 2002.

\bibitem{Andersen03}
L.~Andersen, J.~Sidenius, and S.~Basu.
\newblock All your hedges in one basket.
\newblock {\em RISK}, 16(11):67--72, 2003.

\bibitem{Artzner99}
P.~Artzner, F.~Delbaen, J.-M. Eber, and D.~Heath.
\newblock Coherent measures of risk.
\newblock {\em Mathematical Finance}, 9:203--228, 1999.

\bibitem{Arvanitis98}
A.~Arvanitis, C.~Browne, J.~Gregory, and R.~Martin.
\newblock A credit risk toolbox.
\newblock {\em RISK}, 11(12):50--55, 1998.

\bibitem{Barco04}
M.~Barco.
\newblock Bringing credit portfolio modelling to maturity.
\newblock {\em RISK}, 17(1):86--90, 2004.

\bibitem{Barndorff91}
O.~E. Barndorff-Nielsen.
\newblock Modified signed log likelihood ratio.
\newblock {\em Biometrika}, 78:557--563, 1991.

\bibitem{Bender78}
C.~M. Bender and S.~A. Orszag.
\newblock {\em Advanced mathematical methods for scientists and engineers}.
\newblock McGraw-Hill, New York, 1978.

\bibitem{Burtschell05}
X.~Burtschell, J.~Gregory, and J.-P. Laurent.
\newblock A comparative analysis of {CDO} pricing models.
\newblock {\em {\tt www.defaultrisk.com}}, 2005.

\bibitem{Daniels87}
H.~E. Daniels.
\newblock Tail probability approximations.
\newblock {\em International Statistical Review}, 55(1):37--48, 1987.

\bibitem{Feuerverger00}
A.~Feuerverger and A.~C.~M. Wong.
\newblock Computation of value-at-risk for nonlinear portfolios.
\newblock {\em J. of Risk}, 3(1):37--55, 2000.

\bibitem{Glasserman05}
P.~Glasserman and J.~Li.
\newblock Importance sampling for portfolio credit risk.
\newblock {\em Management Sci.}, 51(11):1643--1656, 2005.

\bibitem{Glasserman06}
P.~Glasserman and J.~Ruiz-Mata.
\newblock Computing the credit loss distribution in the gaussian copula model:
  a comparison of methods.
\newblock {\em J. Credit Risk}, 2(4), 2006.

\bibitem{Gordy02}
M.~B. Gordy.
\newblock Saddlepoint approximation of {C}redit{R}isk+.
\newblock {\em J. Banking and Finance}, 26(7):1337--1355, 1998.

\bibitem{Gordy03a}
M.~B. Gordy.
\newblock A risk-factor model foundation for ratings-based bank capital rules.
\newblock {\em J. Financial Intermediation}, 12(3):199--232, 2003.

\bibitem{Lehrbass04}
M.~Gundlach and F.~Lehrbass (eds).
\newblock {\em CreditRisk+ in the Banking Industry}.
\newblock Springer, 2004.

\bibitem{Jensen95}
J.~L. Jensen.
\newblock {\em Saddlepoint Approximations}.
\newblock Oxford (Clarendon Press), 1995.

\bibitem{Kolassa94}
J.~E. Kolassa.
\newblock {\em Series Approximation Methods in Statistics}.
\newblock Springer, 1994.

\bibitem{Martin98a}
R.~J. Martin.
\newblock System failure statistics: Some asymptotic formulae.
\newblock {\em GEC J. Technology}, 15(1):10--15, 1998.

\bibitem{CSFB04}
R.~J. Martin.
\newblock {\em Credit Portfolio Modeling Handbook}.
\newblock Credit Suisse First Boston, 2004.

\bibitem{Martin06b}
R.~J. Martin.
\newblock The saddlepoint method and portfolio optionalities.
\newblock {\em RISK}, 19(12):93--95, 2006.

\bibitem{Martin06a}
R.~J. Martin and R.~Ordov\'as.
\newblock An indirect view from the saddle.
\newblock {\em RISK}, 19(10):94--99, 2006.

\bibitem{Martin07a}
R.~J. Martin and D.~Tasche.
\newblock Shortfall: {A} tail of two parts.
\newblock {\em RISK}, 20(2):84--89, 2007.

\bibitem{Martin01b}
R.~J. Martin, K.~E. Thompson, and C.~J. Browne.
\newblock Taking to the saddle.
\newblock {\em RISK}, 14(6):91--94, 2001.

\bibitem{Martin01d}
R.~J. Martin, K.~E. Thompson, and C.~J. Browne.
\newblock Va{R}: {W}ho contributes and how much?
\newblock {\em RISK}, 14(8):99--102, 2001.

\bibitem{Martin02a}
R.~J. Martin and T.~S. Wilde.
\newblock Unsystematic credit risk.
\newblock {\em RISK}, 15(11):123--128, 2002.

\bibitem{NRC}
W.~H. Press, B.~P. Flannery, S.~A. Teukolsky, and W.~T. Vetterling.
\newblock {\em Numerical Recipes in C++}.
\newblock CUP, 2002.

\bibitem{Sato02}
K.-I. Saito.
\newblock {\em L\'evy Processes and Infinitely Divisible Distributions}.
\newblock CUP, 2002.

\bibitem{Takano08}
Y.~Takano and J.~Hashiba.
\newblock A novel methodology for credit portfolio analysis: {N}umerical
  approximation approach.
\newblock {\em {\tt www.defaultrisk.com}}, 2008.

\bibitem{ThompsonK03a}
K.~E. Thompson and R.~Ordov\'as.
\newblock Credit ensembles.
\newblock {\em RISK}, 16(4):67--72, 2003.

\bibitem{ThompsonK03b}
K.~E. Thompson and R.~Ordov\'as.
\newblock The road to partition.
\newblock {\em RISK}, 16(5):93--97, 2003.

\bibitem{Huang07}
X.~Wang and C.~Oosterlee.
\newblock Computation of {V}a{R} and {V}a{R} contribution in the {V}asicek
  portfolio credit loss model: a comparative study.
\newblock {\em J. Credit Risk}, 3(3), 2007.

\end{thebibliography}

\end{document}